\documentclass{article}

\usepackage{graphicx} 
\usepackage{jcappub}
\usepackage{xcolor}
\usepackage{subfig}  
\usepackage{placeins}
\usepackage{stmaryrd}
\usepackage{cleveref}
\usepackage{fontawesome5}
\usepackage{hyperref}

\newcommand{\nobs}{N_\mathrm{obs}}

\newcommand{\Pc}[2]{ P \left(#1 \,\middle| \, #2\right)}

\newcommand{\PCc}[2]{\mathring{P}\left(#1 \,\middle| \, #2\right)}

\renewcommand{\d}{\mathrm{d}}

\newcommand{\scalarprod}[2]{\left( #1 \, \left|\, #2\right.\right)}

\newcommand{\Vect}{\boldsymbol{\theta}}
\newcommand{\VecT}{\boldsymbol{\Theta}}
\newcommand{\github}[1]{%
   \href{#1}{\faGithubSquare}%
}
\newcommand{\wfD}{\texttt{IMRPhenomD}}
\newcommand{\wfHM}{\texttt{IMRPhenomHM}}

\newcommand{\gwjulia}{\texttt{GWInference}}

\title{Tests of General Relativity with Einstein Telescope}
\author[a,b,*]{Andrea Begnoni,}
\author[c,d]{Walter Del Pozzo,}
\author[a,b,e,*]{Matteo Pegorin,}
\author[c,d,*]{Joachim Pomper,}
\author[c,d]{Angelo Ricciardone}

\affiliation[a]{Dipartimento di Fisica e Astronomia ``Galileo Galilei'', Universit{\`a} degli Studi di Padova, via Marzolo 8, I-35131 Padova, Italy}
\affiliation[b]{INFN, Sezione di Padova, Via Marzolo 8, I-35131 Padova, Italy}
\affiliation[c]{Dipartimento di Fisica “Enrico Fermi”, Università di Pisa, Largo Bruno Pontecorvo 3, Pisa I-56127, Italy}
\affiliation[d]{INFN, Sezione di Pisa, Largo Bruno Pontecorvo 3, Pisa I-56127, Italy}
\affiliation[e]{Max Planck Institute for Gravitational Physics (Albert Einstein Institute), Am M{\"u}hlenberg 1, Potsdam 14476, Germany}
\note[*]{Corresponding authors.}
\emailAdd{andrea.begnoni@phd.unipd.it}
\emailAdd{walter.delpozzo@unipi.it}
\emailAdd{matteo.pegorin.1@phd.unipd.it}
\emailAdd{joachim.pomper@phd.unipi.it}
\emailAdd{angelo.ricciardone@unipi.it}

\abstract{Gravitational wave signals from compact binary coalescences provide powerful and reliable probes of General Relativity. To date, the LIGO-Virgo-KAGRA collaboration has provided stringent consistency tests of General Relativity predictions.
In this work, we present forecasts for how accurately General Relativity can be tested using third-generation ground-based interferometers, focusing on Einstein Telescope (ET) and binary black hole mergers. 
Given the expected high detection rate, we use a Fisher matrix approach to simulate parameter estimation over large source catalogs, and jointly analyze events within a Bayesian hierarchical framework.
We compare different ET configurations and forecast constraints on parametrized deviations of the inspiral post-Newtonian coefficients. 
Our results indicate that, within about four months of observations, ET could improve current LVK GWTC-3 bounds by roughly four orders of magnitude for the dipole radiation term and about two orders for higher-order coefficients.
We further assess the number of detections required to confidently identify deviations from General Relativity at various post-Newtonian orders and for different detector configurations.}
    

\begin{document}

\maketitle

\pagebreak

\section{Introduction}
    After a decade of observations \cite{LIGOScientific:2016_150914, LIGOScientific:2017MM, LIGOScientific:2018mvr, LIGOScientific:2020ibl, LIGOScientific:2021usb, KAGRA:2021TC3, LIGOScientific:2025slb}, gravitational waves (GWs) from compact binary coalescence (CBCs) have emerged as a powerful tool for testing General Relativity (GR) in the strong-field regime, and highly relativistic regime.
There are several well motivated reasons to consider that GR is not the full theory of gravity, especially when trying to reconcile classical General Relativity with the fundamental principles of quantum mechanics.
Therefore, many beyond-GR theories have been proposed and numerous works have attempted to search for GR deviations in the GW data, observed by the LIGO-Virgo-KAGRA (LVK) collaboration \cite{LIGOScientific:2019fpa, LIGOScientific:2020tif, LIGOScientific:2021test, Sanger:2024axs, LIGOScientific:2025rid, LIGOScientific:2025csr, LIGOScientific:2025brd}. These studies face the difficulty posed by the large number of ways GR deviations could manifest \cite{Berti:2015itd, Ishak:2018his, Payne:2024yhk,Bernard:2025dyh} and the experimental challenges in detecting these deviations \cite{Will:2014kxa}. 
Moreover, this search is also affected by the presence of possible systematics, non-stationary or non-Gaussian noise \cite{Kwok:2021zny, Reali:2023eug}, environmental effects \cite{Roy:2024rhe, Gupta:2024gun, Maggio:2022hre, Pompili:2025cdc}, waveform biases \cite{Brown:2024joy, Dhani:2024jja} or parameter degeneracy and sampling issues \cite{Sanger:2024axs}, which could mimic a deviation from GR.  Extensive studies have also scrutinized the nature and close environment of compact objects \cite{Cardoso:2019rvt, Lunin:2001jy, Mazur:2004fk, Liebling:2012fv, Giudice:2016zpa, Danielsson:2021ykm, Almheiri:2012rt, Brito:2015oca, Cardoso:2016oxy, Cardoso:2016rao, Cardoso:2017cqb, Maggio:2021ans, Chakraborty:2022zlq, Silvestrini:2025lbe, Mastrogiovanni:2025ixe, Berti:2025hly, Ianniccari:2024ysv, Forteza:2022tgq, Baumann:2018vus, Baumann:2022pkl, Baumann:2019eav, Ikeda:2020xvt, Baumann:2021fkf,  DellaMonica:2025zby, Barsanti:2022vvl, Maselli:2023khq, DellaRocca:2024sda, Bambi:2025wjx, Begnoni:2025aqc, Nicolini:2025yhu, Kejriwal:2025jao}. To date, no GR deviation has been detected by the LVK collaboration \cite{LIGOScientific:2021test, Sanger:2024axs}.

In this work, we adopt a theory-agnostic approach, first developed in \cite{Agathos:2013TIGER} and employed in the LVK’s TIGER pipeline, to perform tests on deviations from GR \cite{Meidam:2017dgf, Li:2011cg}. The spirit of this approach, as used in the LVK analysis \cite{LIGOScientific:2021test, LIGOScientific:2026fcf}, is to add parametrized GR deviations in the post-Newtonian (PN) coefficients \cite{Blanchet:2013haa, Thorne:1980ru} and perform inference by varying one coefficient at a time. These modifications are applied to the inspiral phase of the waveform, followed by adjustments to the phase in the merger and ringdown parts to guarantee overall $\mathcal{C}^1$-regularity in the frequency domain \cite{Khan:2015PhD}. 
Another more recent framework to test GR in the inspiral phase is the Flexible Theory Independent method \cite{Mehta:2022pcn,Sanger:2024axs,Piarulli:2025rvr}, which uses a taper function to isolate the inspiral deviation, reducing leakage to the merger and ringdown phases. 
Additional methods for testing deviations from GR are also the parameterized post-Newtonian \cite{Nordtvedt:1968qs, Will:1972zz, Will:1971zzb, Nordtvedt:1972zz, Will:1973zz, Will:2014kxa}, the parameterized post-Keplerian \cite{Damour:1990wz, Damour:1991rd} and the post-Einsteinian \cite{Yunes:2009ke, Cornish:2011ys, Chatziioannou:2012rf, Yunes:2016jcc, Tahura:2018zuq, Bonilla:2022dyt, Loutrel:2022xok, Xie:2024ubm, Xie:2025voe} frameworks.

Many beyond GR theories motivate an accurate scrutiny of the PN terms, e.g., higher curvature theories \cite{Endlich:2017tqa, Sennett:2019bpc, Bernard:2025dyh}, Dynamical Chern-Simons \cite{Yunes:2009hc,Molina:2010fb}, Einstein-Maxwell-Dilaton \cite{Julie:2017rpw, Khalil:2018aaj,Julie:2018lfp} or Einstein-Dilaton-Gauss-Bonnet \cite{Kanti:1995vq, Maselli:2015tta, Blazquez-Salcedo:2016enn, Sennett:2017lcx, Silva:2017uqg, Witek:2018dmd, Kuntz:2019zef, Khalil:2019wyy, Julie:2019sab, Perkins:2021mhb, Julie:2022huo, Bernard:2022noq, Julie:2022qux, Creci:2023cfx, Bernard:2023eul, Diedrichs:2023foj, Almeida:2024uph, Julie:2024fwy, Almeida:2024cqz, Dones:2025zbs}, for a review \cite{Yunes:2013dva}, and GWs represent a promising tool to probe such GR deviations \cite{LIGOScientific:2016lio, LIGOScientific:2018dkp, LIGOScientific:2019fpa, LIGOScientific:2017ycc, LIGOScientific:2021test, LIGOScientific:2020tif, Chamberlain:2017fjl, Perkins:2020tra, Gupta:2020lxa}. 
To improve the constraints on potential GR deviations, 
we can take advantage of the large number of GW detections expected from the third generation of interferometers. For this reason, we forecast the improvements achievable by hierarchically combining information from many individual GW events, 
utilizing a Gaussian population model \cite{Isi_2019}. 
We show that even small deviations from GR can be probed with Einstein Telescope (ET) \cite{ET:2019Magg, ET:2025xjr}.

ET is a third generation GW interferometer \cite{Punturo:2010zza}, expected to improve the detector sensitivity with respect to second generation interferometers and to extend the observable horizon; it has been included in the 2021 European Strategy Forum on Research Infrastructures (ESFRI) road map. The location and the final layout are still under discussion in the community \cite{Branchesi:2023COBA, ET:2025xjr}: possible layouts are a triangular configuration ($\Delta$) or a 2L shape detector network (\texttt{2L}) with the arms either aligned (\texttt{2L\_0}) or misaligned by 45 degrees (\texttt{2L\_45}). In this work, we compare such different ET designs to estimate their forecasting power to measure GR deviations. 
To perform the forecast, we employ a Fisher matrix approach to analyze a large catalog of sources with complete inspiral-merger-ringdown (IMR) waveforms, in particular, the Phenom-family waveforms \cite{London:2017HM, Kalaghatgi:2019HM, Pratten:2020XAS, Husa:2015PhD, Khan:2015PhD}. 

We use waveform models that include not only the fundamental mode but also higher harmonics \cite{London:2017HM, Kalaghatgi:2019HM}, highlighting the importance of higher-mode waveforms for testing deviations from GR, and focusing on models that incorporate spin effects only for aligned spins.
The analysis focuses on Binary Black-Holes (BBH) and involves hundreds of thousands of individual GW events. It is carried out using the \gwjulia code \cite{Begnoni:2025oyd}, which allows the evaluation of the individual Fisher matrix, for all GR and beyond-GR waveforms, in a very fast and efficient manner.

The paper is organized as follows. In Sec.~\ref{sec:methodoloy} we describe the theoretical and computational methodology employed in this work; in Sec.~\ref{sec:verification_with_lvk} we validate our Fisher-matrix forecasts against observational results from the LVK collaboration~\cite{LIGOScientific:2021test}. Section~\ref{sec:forecast_et_gr_correct} presents forecasts for ET’s ability to test GR assuming GR as the fiducial theory, whereas Sec.~\ref{sec:forecast_et_gr_wrong} evaluates ET’s constraining capabilities when assuming deviations from GR. The discussion and conclusions are presented in Sec.~\ref{sec:conclusion}. App.~\ref{app:catalog_detector} details the catalog and detector setup, App.~\ref{app:f_min} examines the impact of the low-frequency cutoff, App.~\ref{app:a3_more_hyperparam_dist} provides further details on the hierarchical hyperparameter distribution within the Fisher approximation, while App.~\ref{app:a4_lvk_additional_checks} examines the impact of higher modes in the waveform model on higher PN order constraints when considering the LVK network. 

    \label{sec:introduction}
    
\section{Methodology}
    \label{sec:methodoloy}
    In this section, we introduce the formalisms and technical definitions used throughout this work, beginning with the analytical modeling of GW, the PN framework and the Fisher information matrix. Then, we give an in-depth explanation of the Bayesian hierarchical methods that will be employed in the following sections.

\subsection{Model independent parametrization of GR deviations}
\label{sec:waveform_model}
    
    A GW signal $h$ in the fundamental mode can be written as 
    \begin{equation}
        h_{+/\times}(t) = \mathcal{A}_{+/\times}(t)\,e^{i\Phi(t)}\,,
    \end{equation}
    where $\mathcal{A}_{+/\times}$ represent the amplitude, $\Phi$ is the phase and $+/\times$ indicate the two GW polarizations. This formulation can be extended to include higher order modes \cite{London:2017HM,Cotesta:2018fcv}. Conventionally, GW waveforms from CBC are divided in inspiral, merger and ringdown phases \cite{maggiore2008gravitational}.
    In particular, the inspiral part has an analytic description within the PN framework, which accurately models the inspiral of comparable-mass compact objects. In this framework, the phase can be written as a Taylor expansion in the characteristic orbital velocity $v$  as $(v/c)^{p}$, where $c$ is the speed of light and $p/2$ indicates the PN order.
    Once one moves to the Fourier domain, within the stationary phase and quasi-circular orbit approximations, the inspiral phase can be expressed as \cite{Cutler:1994ys,Damour:2000zb, Buonanno:2009F2}
    \begin{equation} \label{eq:pn_inspiral_phase_formula}
    \Phi(f)=2 \pi f t_c-\Phi_c-\frac{\pi}{4}+\sum_{p=0}^7\left[\varphi_p+\varphi_{p\,l} \ln f\right] f^{(p-5) / 3}\,,
    \end{equation}
    where $\varphi_p$ indicates the PN term coefficient, $f$ indicates the frequency, $t_c$ and $\Phi_c$ the time and phase of coalescence of the binary, respectively. All the PN terms $\varphi_p$ with $p\in \{0,\dots,7\}$, as well as $\varphi_{5l}$ and $\varphi_{6l}$ have a known analytic expression, e.g., \cite{Buonanno:2009F2,Blanchet:2023bwj}.
    Each PN term $\varphi_p$ in GR is divided into spin-dependent and spin-independent components
    \begin{equation}
        \varphi_p^{\rm{GR}} = \varphi_p^{\rm{GR,\, NS}} + \varphi_p^{\rm{GR,\, S}}\,.
    \end{equation}
    In this work, we consider modifications to the phase $\Phi$ in the inspiral part of the coalescence, due to beyond GR effects.
    The deviations to the phase are parametrized by adding the deviation coefficients $\delta\varphi_p$ to each PN coefficient $\varphi_p$, where the index $p$ now includes also the log terms. Perturbations may be added to the spin-dependent and spin-independent parts of the PN coefficients separately
    \begin{equation}
        \delta\varphi_p = \delta\varphi_p^{\rm NS} +  \delta\varphi_p^{\rm S} \,.
    \end{equation}
    We follow the procedure described in \cite{LIGOScientific:2020tif}; thus, we introduce modifications only to the spin-independent part, leaving the spin-dependent part unaltered, i.e., we assume $ \delta\varphi_p^{\rm S} = 0$. The modification of the PN coefficients then amounts to 
    \begin{equation}
         \varphi_p^{\rm } =\left(1 + \delta\varphi_p \right)\varphi_p^{\rm{GR,\,NS}} + \varphi_p^{\rm{GR,\,S}}\;.
    \end{equation}
    This implementation as a relative deviation is valid when $\varphi_p^{\rm{GR,\,NS}} \neq0 $. In the other cases $\delta\varphi_p$ represent an absolute deviation from zero. 
    No deviation is introduced for the spin-induced quadrupole moment.
    
    We implement these modifications to two waveform models calibrated to numerical relativity: \wfD~\cite{Husa:2015PhD, Khan:2015PhD} and \wfHM~\cite{London:2017HM, Kalaghatgi:2019HM}.
    The main difference between the two waveforms is that \wfHM~contains higher-order modes, with respect to the fundamental (2,2) mode, being the only mode present in \wfD. These higher-order modes are key for breaking the degeneracy between the inclination angle and the luminosity distance, e.g., \cite{Iacovelli:2022For}, and allow for an overall better parameter estimation 
    \cite{Roy:2025gzv, Christensen:2022bxb}.
    The waveform models \wfD~and \wfHM~assume aligned spins, thus neglecting precession effects.  In principle, precession can be taken into account using a twisting-up technique \cite{Roy:2025gzv}, but has been neglected for the Fisher analysis presented here.
    After the modification of the inspiral parameter in \eqref{eq:pn_inspiral_phase_formula}, phase- and time-shift parameters of the waveform template at the boundaries between the different regimes of the model are adjusted, such that $\mathcal{C}^1$-continuity of the waveform is maintained in the frequency domain. 
    The modifications to the inspiral part implemented in the \wfD~and \wfHM~waveform models do not significantly affect the intermediate and merger-ringdown parts\footnote{We did not employ waveforms of the \texttt{PhenomX} family since the implementation of the deformation coefficients is more subtle \cite{Roy:2025gzv}.}.
    
    Following the TIGER approach \cite{Agathos:2013TIGER}, in the analysis we add a single deviation parameters at a time, while fixing the others to zero. This procedure provides a modified inspiral-merger-ringdown (IMR) waveform model:
    \begin{align}
        h_{\varphi_p}[\VecT](f) &= h^{\varphi_p}_+[\VecT] F_+ + h^{\varphi_p}_\times[\VecT] F_\times \;,
    \end{align}
    where $F_{+/\times}$ denotes the antenna pattern functions of the detector and $h^{\varphi_p}_{+/\times}$ are the individual polarization contributions for which a deviation parameter was added to the PN coefficient $\varphi_p$. 
    We combined the deviation parameter $\delta \varphi_p$ with the other standard waveform parameter $\Vect$ in an extended vector $\VecT = (\Vect, \delta \varphi_p)$. Thus, as we are neglecting precession effects, our analysis considers 11 GR waveform parameters, reducing the information of the black-hole spins to only the component normal to the plane of motion.
    Hence, in this work, $\VecT$ is a vector of the total 12 parameters, see App.~\ref{app:catalog_detector} for more information. 
    Considering the following set of potential deviations, this provides 10 IMR waveforms, one for each selected deviation.
    \begin{equation}
        \delta\varphi_p \in \left\{\delta\varphi_{-2}, \delta\varphi_{0}, \delta\varphi_{1}, \delta\varphi_{2}, \delta\varphi_{3},
        \delta\varphi_{4}, \delta\varphi_{5l}, \delta\varphi_{6}, \delta\varphi_{6l}, \delta\varphi_{7}\right\}
    \end{equation}
    We are not considering $\varphi_5$, since this parameter is degenerate with $\Phi_c$ for non-precessing models, evident from \eqref{eq:pn_inspiral_phase_formula}. Furthermore, the PN coefficient for $p = -2$ and $p = 1$ vanish in GR and thus $\delta\varphi_{-2}$ and $\delta\varphi_1$ are parametrized as total deviation defined as
    
    \begin{equation}
    \Phi^{p}_{\ell m} =\frac{3}{128 \eta}\tilde f^{(p-5) / 3} \delta \varphi_p\,,
    \end{equation}
    where $\tilde f = 2 \pi M (1+z) f / m$, with $M(1+z)$ the total mass at detector, $\eta$ the symmetric mass ratio, $z$ the redshift and $(\ell, m)$ indicate the spherical harmonics multipoles.  
    A non-vanishing $\delta\varphi_{-2}$ coefficient is typically associated with leading-order dipole radiation, which can arise in specific modified theories of gravity that evade the black hole no-hair theorem, such as Einstein-dilaton-Gauss-Bonnet gravity for binaries with asymmetric scalar charges \cite{Israel:1967wq, Israel:1967za, Carter:1971zc, Hawking:1971vc, Hawking:1972qk, Robinson:1975bv, Bekenstein:1995un, Sotiriou:2011dz, Barausse:2016eii, Silva:2017uqg, Antoniou:2017acq, Doneva:2017bvd, Witek:2018dmd, Silva:2018qhn, Khalil:2019wyy, Julie:2019sab}. 
    Moreover, among the PN deviation coefficients inferred from direct GW observations, the dipole term, $\delta\varphi_{-2}$, is expected to be the most tightly constrained.
    Higher PN terms become relevant later in the inspiral, due to their dependence on higher powers of the orbital velocity, as evident from  \eqref{eq:pn_inspiral_phase_formula}. 
    This has relevant consequences, e.g., low mass events result in the best constraints for the low PN terms, due to the large number of cycles observed during their early inspiral phase. Another important aspect is that the low PN orders depend significantly on the low frequency limit $f_{\rm min}$ of the analysis, which will be lower in next generation detectors, hence resulting in significant improvements, as shown in App. \ref{app:f_min}. 
    
     Assuming stationarity and Gaussianity, we describe the detector noise using the (one-sided) power-spectral-density (PSD) in Fourier space $S_n(f)$. Using the noise weighted scalar product of two functions $a,b$ in the frequency domain
    \begin{equation}
        \scalarprod{a}{b} := 2 \int_{f_\mathrm{min}}^{f_\mathrm{max}} \frac{a(f) b^*(f)+a^*(f) b(f)}{S_n(f)} \mathrm{d} f\,,
    \end{equation}
    we introduce the optimal signal-to-noise ratio (SNR) of a signal template in a given detector as 
    \begin{equation} \label{eq:definition_optimal_snr}
        {\rm SNR}[\VecT] := \sqrt{ \scalarprod{h[\VecT]}{h[\VecT]}}\ .
    \end{equation}
    We can then introduce the SNR of the network of detectors as \begin{equation}
        \text{SNR}_\mathrm{Network}=\sqrt{\sum_d \text{SNR}_d^2}\,,
    \end{equation}
    where the sum is performed over the SNR of the individual detector labeled by $d$. Note that there are three detectors for a triangular design, which in our treatment are considered to be statistically independent, i.e., the joint likelihood is the product of the three detector likelihoods \cite{Freise:2008dk}. 
    In the following, we will approximate the single event posteriors by an expansion in the combined detector SNR, obtaining the Fisher Information Matrix. 
    
\subsection{Hierarchical analysis with Fisher approximation}
\label{sec:hierarchical_analysis}

In order to estimate the capability of a detector network to constrain the deviations in the PN coefficients $\delta\varphi_p$, we use a Bayesian hierarchical framework to combine the information $D^{\nobs} \equiv \bigcup_{k=1}^{\nobs} D_k$ of $\nobs$ individual gravitational wave observations with datasets $D_k$, into an overall posterior for the PN coefficients. These individual events originate from a presumably unknown population of black hole binaries.
Following \cite{Isi_2019}, we adopt a minimal information approach, modeling the population distribution of $\delta\varphi_{p}$ as a Gaussian $\mathcal{N}\left(\delta \varphi_{p}\,\vert\,\mu,\sigma\right)$ with two population hyperparameters $\mu$ and $\sigma$. As discussed earlier, we consider deviations in PN orders one at a time and the population hyperparameters in principle may differ between the PN orders under consideration. The posterior of $\delta\varphi_{p}$ can be written as
\begin{equation} \label{eq:deviation_dist}
    \Pc{\delta \varphi_{p}}{D^{\nobs}, I} = \int \d \mu\d\sigma\; \mathcal{N}\left(\delta \varphi_{p}\,\vert\,\mu,\sigma\right)\Pc{\mu,\sigma}{D^{\nobs}, I} \,.
\end{equation}
Here, $I$ denotes our general information background in the Bayesian sense. 
The population posterior $\Pc{\mu,\sigma}{D^{\nobs}, I}$ of the hyperparameters is the object of interest, combining the information of $\nobs$ individual GW observations. Using Bayes’ theorem and assuming that the observations are independent, we obtain
\begin{equation} \label{eq:population_distribution_inference}
    \Pc{\mu, \sigma}{D^{\nobs}, I}  = \frac{\Pc{\mu, \sigma}{I}}{\Pc{D^{\nobs}}{I}} \prod_{k=1}^{\nobs} \int \d\delta\varphi_p\; \Pc{D_k}{\delta\varphi_p, I} \mathcal{N}(\delta\varphi_p\,\vert\,\mu, \sigma)\,.
\end{equation}
The posterior \eqref{eq:population_distribution_inference} is computed
by marginalizing over the individual events, as 
\begin{equation} \label{eq:marginalized_single_event_pe_posterions}
    \Pc{D_k}{\delta\varphi_p, I} = \int \d \Vect \; \Pc{D_k}{\Vect, \delta\varphi_p, I}\Pc{\Vect}{\delta \varphi_p, I} \,.
\end{equation}
For forecasting purposes, we apply a linear signal approximation (LSA) to the likelihoods \cite{Vallisneri_2008} of the individual events, labeled by an index $k= 1,\dots,\nobs$.
Such an approximation may be justified by the high SNR of signals detected by next-generation detectors, such as ET. To do so, one expands the likelihood in inverse powers of the optimal 
SNR of the true signal $ h[\VecT^0_k]$ (here $\VecT^0_k$ denotes the fiducial point in parameter space, characterizing the $k$-th observed event) and then self-consistently truncates the perturbative series, keeping only the lowest order terms \cite{Vallisneri_2008}. For each event, labeled by the index $k$, the expansion can be written as 
\begin{align} \label{eq:lsa_likelihood}
    \ln \Pc{D_k}{\VecT, I} = -\frac{1}{2}(\VecT - \overline{\VecT}_k)^\top \Gamma_k \left(\VecT - \overline{\VecT}_k\right)
    + \; \mathcal{O}\left(\frac{1}{\mathrm{SNR}[\VecT_k^0]}\right)\,,
\end{align} 
where we omit terms independent of $\VecT$, since they can be absorbed in the normalization. 
After truncation at the lowest order, the resulting likelihood is completely determined by the Fisher information matrix
\begin{align}
    \left\{\Gamma_k\right\}_{i j} = \scalarprod{\vec{\nabla}_{x_i} h[x_1, x_2, \dots]}{  \vec{\nabla}_{x_j} h[x_1, x_2, \dots]} \Big \vert_{\mathbf{x} = \VecT^{0}_k} 
\end{align}
and the maximum of the approximated likelihood
\begin{equation}\label{eq:likelihood_peak}
     \left\{\overline{\VecT}_k\right\}_i = \sum_{j} \left\{\Gamma_k^{-1}\right\}_{ij} \scalarprod{n_k}{\vec{\nabla}_{x_j} h[x_1, x_2, \dots]} \Big \vert_{\mathbf{x} = \VecT^{0}_k}\,.
\end{equation}
In the last expression, $n_k$ denotes an explicit noise realization of the detector. Hence, $\overline{\VecT}_k$ is a random variable and depends on the noise realization, while the Fisher matrix is independent of the noise. 
When combining multiple events within the hierarchical framework, the overlap, and thus the center, of the individual likelihoods is of great importance \cite{pacilio_catalog_2024}. In fact, in the limit of large $\nobs$, the results turn out to be heavily biased when the Fisher-approximated posteriors are always centered around the values $\VecT_k^0$. This highlights the importance of correctly accounting for the noise-dependent shift of the maximum of the likelihoods.
For technical implementations, we will only need to calculate the Fisher matrices $\Gamma_k$ of the individual events, since in the LSA the integrals in \eqref{eq:population_distribution_inference} and \eqref{eq:marginalized_single_event_pe_posterions} can be performed analytically. In the following, we will assume
\begin{equation} \label{eq:deviation_prior_distribution}
    \Pc{\Vect}{\delta \varphi_p,I} = \Pi\left(\Vect \vert I\right) = \mathrm{const}.
\end{equation} 
While the independence of the distribution of potentially known deviations could be physically reasonable, we stress that the assumption of an entirely flat single event prior $ \Pi\left(\Vect \vert I\right)$ is purely a technical tool and certainly does not reflect what is done in a full Bayesian analysis. Since we are concerned with combining a large number of informative events, we can also put a flat prior $\Pc{\mu, \sigma}{I}$ for $\mu$ and $\sigma$. 
In the limit of large $\nobs$ we can safely assume the resulting posterior to be likelihood driven. If we define
\begin{equation}\label{eq:delta_k_definition}
    \Delta_k^2 := \left\{\Gamma^{-1}_k\right\}_{\delta\varphi_p\delta\varphi_p} \quad\text{and}\quad \delta\overline{\phi}_k := \left\{\overline{\Theta}_{k}\right\}_{\delta\varphi_p} \;,
\end{equation}
we then obtain the posterior distribution for the two population hyperparameters $\mu$ and $\sigma$ \cite{zhong_multidimensional_2024}
\begin{align} \label{eq:posterior_hyperparamter_tiger}
    \Pc{\mu, \sigma}{D^{\nobs},I} \propto
    \prod_{k=1}^{\nobs} \sqrt{\frac{\Delta_k^2}{\Delta_k^2 + \sigma^2}}\exp\left({-\frac{1}{2}} \frac{(\mu-\delta \overline{\phi}_{k})^2}{\Delta_k^2 + \sigma^2}\right) \;,
\end{align}
which was the initial object of interest \eqref{eq:population_distribution_inference}, encoding all the relevant information about the distribution of the PN deformation coefficients $\delta\varphi$, under the minimal information assumption.
The quantities $\delta\overline{\phi}_{k}$ can be shown to behave as Gaussian random variables 
\begin{equation}
    \delta\overline{\phi}_{k} = \left\{\VecT^0_{k}\right\}_{\delta\varphi_p} + \Delta_k \,\mathcal{X}\,,
    \label{eq:delta_phi_noise_realization}
\end{equation}
with $\mathcal{X} \sim \mathcal{N}(0,1)$ a unit Gaussian random variable \cite{Vallisneri_2008}. Having computed the Fisher matrix, we can account for different detector noise realizations by sampling from a standard normal and adding the corresponding shifts to the fiducial value. 

The hierarchical distribution includes certain limiting scenarios of the population distribution, such as the limit of $\sigma \to 0$. For this particular subspace in the population parameter space, the PN deformation coefficient $\delta\varphi_p$ takes on the same value $\mu$ for all individual events\footnote{$\mu$ is still a function of PN order. We omit the index $p$ in the notation. }. 

From the hierarchical distribution, we may obtain the distribution of the deviations via \eqref{eq:deviation_dist}. 
Under the assumption of sufficiently wide priors, the integral over $\mu$ can again be performed analytically, while the integral over $\sigma$ must be evaluated numerically. We obtain
\begin{equation}\label{eq:deviation_dist_result}
    \Pc{\delta \varphi_p}{D^{\nobs},I} \propto \int_0^\infty \mathrm{d}\sigma \frac{e^{c(\sigma)}}{\sqrt{1+a(\sigma)\sigma^2}}\exp{\left(-\frac{1}{2}\frac{a(\sigma)\delta\varphi^2_p  - 2 b(\sigma)\delta\varphi_p - b^2(\sigma)\sigma^2}{1+a(\sigma)\sigma^2}\right)}\;,
\end{equation}
where we define the auxiliary functions
\begin{align}
    a(\sigma) &= \sum_{k=1}^{\nobs} \frac{1}{\Delta_k^2 + \sigma^2}\,,
    \quad\quad b(\sigma) = \sum_{k=1}^{\nobs} \frac{\delta \overline{\phi}_k}{\Delta_k^2 + \sigma^2}\,,\\
    c(\sigma) &= \frac{1}{2}\sum_{k=1}^{\nobs} \left( \ln\left(\frac{\Delta_k^2}{\Delta_k^2 + \sigma^2}\right) - \frac{\delta \overline{\phi}^2_k}{\Delta_k^2 + \sigma^2} \right)  \,.
\end{align}
For the Fisher approximation, we need to fix a fiducial point around which the series expansion is performed. For the forecast of consistency tests assuming GR as the null hypothesis, $\delta \varphi_p^0 = 0$ is the fiducial point for every event of the catalog. In this special case, the null-hypothesis is contained within the $\sigma \to 0$ subspace of the population parameter space. We thus consider also the 
deviation distribution
\begin{align}
    \PCc{\delta\varphi_p}{D^{\nobs}, I} &= \lim_{\sigma \to 0} \int \d \mu\; \mathcal{N}\left(\delta \varphi_{p}\,\vert\,\mu,\sigma\right)\Pc{\mu,\sigma}{D^{\nobs}, I} \\
    \label{eq:conditioned_deviation_posterior}
    &\propto \prod_{k=1}^{\nobs} \exp\left(-\frac{1}{2}\frac{(\delta \varphi_p - \delta \overline{\phi}_k)^2}{\Delta_k^2}\right)
\end{align}
conditioned to the hyperparameter subspace, for which the deviation manifests the same among all individual events, by replacing the flat prior on $\sigma$ with a $\delta$-function conditioned on $\sigma = 0$. In this case, the combined deviation distribution reduces to the product of the individual deviation distributions.
Clearly, the conditioned distribution $\PCc{\delta\varphi}{D^{\nobs}, I}$ is always narrower than the full hierarchical deviation distribution \eqref{eq:deviation_dist_result}. We refer to Appendix~\ref{app:a3_more_hyperparam_dist} for a further comparison. 
Thus, combined with the omission of the information in the tails of the posterior due to the Fisher-approximation, we use the conditioned distribution $\PCc{\delta\varphi}{D^{\nobs}, I}$ to estimate lower bounds of our capability to constrain deviations in the PN coefficients with future detectors.
To do so, we calculate the $90\%$ credible upper bound of the deviation posterior $\PCc{\delta\varphi_p }{D^{\nobs}, I}$, defined as the absolute value of the PN deviation coefficient
\begin{equation}\label{eq:90ci_upper_bounds_deformation_coefficients}
    |\delta\varphi_p|\quad \mathrm{such\;that}\quad \int_{-|\delta\varphi_p|}^{+|\delta\varphi_p|} \PCc{\delta\varphi_p }{D^{\nobs}, I} = 0.9\quad 
\end{equation}
holds. 
        
\section{Validation of the setup with LVK}
    To validate our full analysis, before applying it to the case of Einstein Telescope, we perform an estimation of the constraining power of a GW observation run, such as the O3b observing run of the LVK collaboration  \cite{LIGOScientific:2021test}.
This is intended to provide verification of our software implementation, as well as for the methods presented in Section~\ref{sec:methodoloy}. 

To do so, we set up a detector network composed of three detectors: LIGO Hanford ($\texttt{H}$), LIGO Livingston ($\texttt{L}$) \cite{LIGOScientific:2014pky}, and Virgo ($\texttt{V}$) \cite{VIRGO:2014yos}. The network composed of these three detectors will be referred to as $\texttt{HLV}$.
We compute the conditioned posterior $\PCc{\delta\varphi_p}{D^{\nobs}, I}$ in \eqref{eq:conditioned_deviation_posterior} from a total of 9 observed events, drawn from a catalog of BBHs and fiducial expansion point $\varphi^0_p=0$. From the combined deviation distribution, we estimate the upper bounds $|\delta\varphi_p|$. For this, an event is defined as observable if it satisfies the following three criteria: i) The event has a network SNR larger than 12, ii) the network SNR contribution from the inspiral is larger than 6, 
iii) the Fisher information matrix of the event is invertible. 
The second criterion has a two-fold purpose. On the one hand, it has been adopted to mimic the selection of events in the analysis by the LVK collaboration \cite{LIGOScientific:2021test}, allowing for a consistent comparison. On the other hand, it guarantees that the signals taken into account in the analysis have a significant amount of SNR in the inspiral portion of the signal. This seems important to avoid potential systematics, given that the parametrization of GR deviation was introduced via phase modifications during the inspiral. The inspiral portion of the signal, used for the second criterion, is defined via the inspiral cutoff frequency $f_{\mathrm{insp,cutoff}}$, defined within the \texttt{Phenom} waveform template family~\cite{Husa:2015PhD}.
The third criterion is purely technical, ensuring the applicability of the Fisher-approximation. The number of events that do not pass the third criterion depends on the waveform and network used; however, it reached at most 1\% of events. 
Contrary to what will be done for ET later, we adopt a frequency cutoff of $f_{\mathrm{min}} = 20$Hz, in order to be able to compare our result with the measurements performed by the LVK. The minimum analysis frequency has a large impact on the final constraints of the PN deviations, in particular at lower PN orders. Further details are discussed in Appendix~\ref{app:f_min}. Details about the generation of the catalog and the used PSDs for the setup of the detectors \texttt{H}, \texttt{L} and \texttt{V} can be found in Appendix~\ref{app:catalog_detector}.

\begin{figure}
    \centering
    \includegraphics[width=1.0\textwidth]{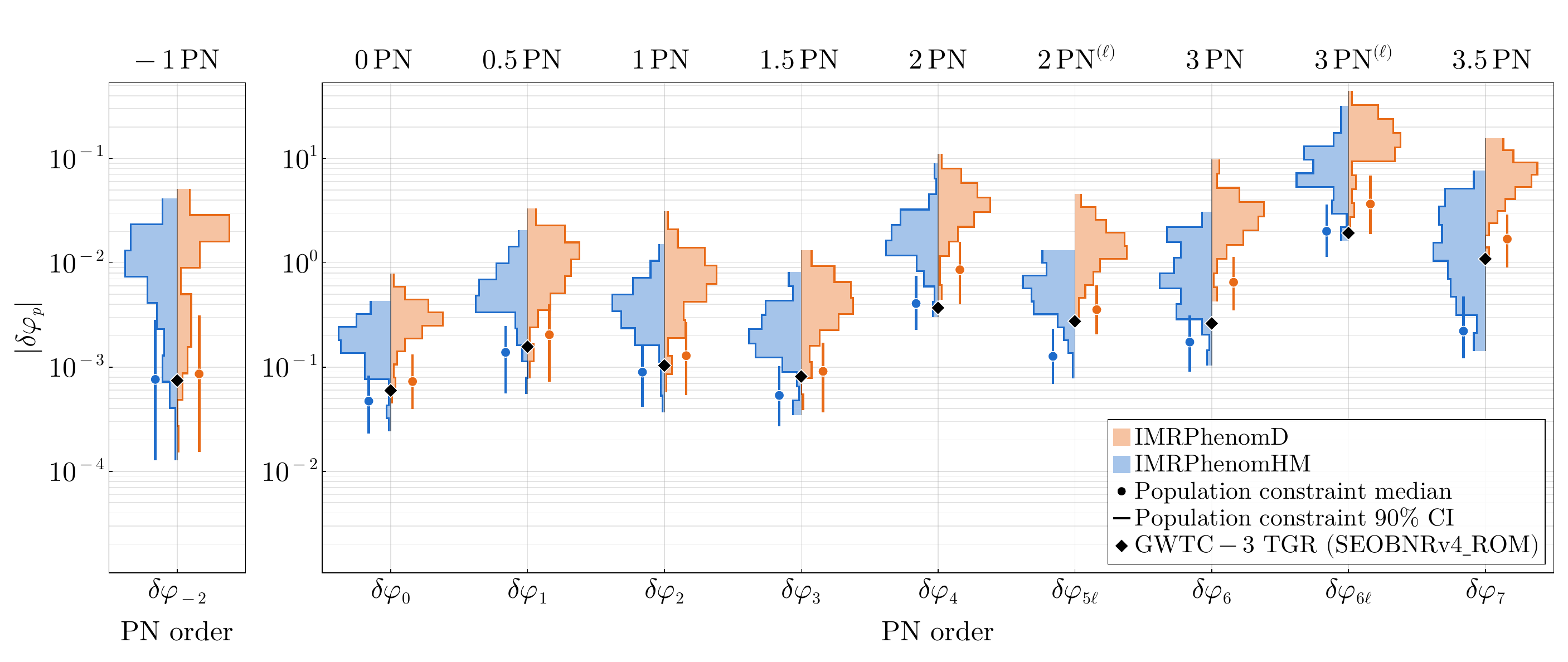}
    \caption{
    Forecast of the $90\%$ upper bounds on the magnitude of the restricted PN deformation coefficients $|\delta\varphi_p|$ for a LVK O3b-like observing run, for the different PN orders considered. The population constraints are obtained from the distribution $\PCc{\delta\varphi_p }{D^{\nobs}, I}$ in \eqref{eq:conditioned_deviation_posterior}, assuming that deviations from GR take on the same value for all observed events.
    The label $\varphi_p$ refers to the $\frac{p}{2}$-PN order coefficient, $\varphi_{p\ell}$ refers to the $\frac{n}{2}$-PN order coefficient with the log term. 
    Each marker and its corresponding error bar represent the median and $90\%$ confidence interval, for $|\delta\varphi_p|$ calculated for $10^3$ independent realizations of the observing run with different noise realizations. 
    The color-coded violin plots show the distribution of the single event $90\%$ upper bounds $|\delta\varphi_p|$.
    We show the results of the forecast employing two different waveform models:  
    \wfHM~and \wfD.
    For comparison, we report the results obtained by the LVK collaboration with GWTC-3, performing a full Bayesian analysis with the \texttt{SEOBNRv4\_ROM} waveform template on the data of $9$ real observed events, see Figure~6 in \cite{LIGOScientific:2021test}. 
    The present FIM analysis predictions are in good agreement with the GWTC-3 results.
    }
    \label{fig:upper_bounds_deformation_coefficients_LVK}
\end{figure}

Using our setup, specified above, allows for a reasonable comparison with LVK's O3b analysis, which employed a full Bayesian analysis applied to the data of 9 real observed events.
Figure~\ref{fig:upper_bounds_deformation_coefficients_LVK} shows a comparison of upper limits on the deviation parameters: our estimates based on \wfD~and \wfHM~compared to the LVK constraints from real data using \texttt{SEOBNRv4\_ROM} \cite{LIGOScientific:2021test}. 
The LVK and \texttt{IMRPhenomD} results, both using only fundamental mode waveforms, are in good agreement, despite the known limitations of the Fisher–matrix approximation.
The colored violin plots in the Figure show the distribution of the $90\%$ upper bounds of $|\delta \varphi_p|$, as obtained from the individual observable events drawn from the catalog. These single-event bounds are evaluated by applying Equations \eqref{eq:conditioned_deviation_posterior} and \eqref{eq:90ci_upper_bounds_deformation_coefficients} to each event separately.
The LVK results quoted represent the upper bounds on PN deviation parameter, obtained with the \texttt{SEOBNRv4\_ROM} waveform template \cite{LIGOScientific:2021test}. For these specific results, also the LVK analysis adopted a $\delta$-prior for $\sigma$, enforcing any GR deviation to manifest with the same value in all individual observations. 
In order to quantify the fluctuations of our estimate of the upper bounds $|\delta\varphi_p|$, 
obtained from the distribution $\PCc{\delta\varphi_p}{D^{\nobs},I}$ in \eqref{eq:deviation_dist_result}, due to noise and the finite catalog realization, we repeat the analysis $10^3$ times.
The fluctuation of the lower bound \cite{pacilio_catalog_2024} is visualized as error bars of the median value of the $10^3$ repetitions, shown in Figure~\ref{fig:upper_bounds_deformation_coefficients_LVK}. The error bars represent the symmetric $90\%$ credible intervals.

A comparison of the forecasts employing \wfD~vs. \wfHM~suggests that the inclusion of higher modes in the analysis results in a $\mathcal{O}(2)$ improvement in the bounds, consistent with results from full Bayesian parameter estimation, e.g., ~\cite{Mehta:2022pcn, Roy:2025gzv}. 
The improvements on the constraints for $\delta\varphi_{5l}$, $\delta\varphi_6$, and $\delta\varphi_7$ exceed the expectations inferred from individual Bayesian analyses showing pronounced tails, which apparently indicate tighter bounds. As discussed in Appendix~\ref{app:a4_lvk_additional_checks}, these tails originate from the contribution of high-mass binary systems, whose signals enter the detector's sensitivity band only during the late inspiral phase. Since the GR deviations were introduced only during the inspiral, loud signals dominated in-band by the post-inspiral phase, may yield overestimated constraints at higher PN orders. Thus, the Fisher forecast for the constraint of these PN coefficients when including higher modes needs to be interpreted with caution.
Moreover, the population mass distribution with a peak at $M_\mathrm{peak} \approx 34M_\odot$ further enhances the prevalence of such high-mass sources. We refer to Appendix~\ref{app:a4_lvk_additional_checks} for a more detailed discussion. 
Nonetheless, it appears that overall the inclusion of higher modes helps to improve constraining deviations from GR \cite{Mehta:2022pcn, Roy:2025gzv}.
    \label{sec:verification_with_lvk}
    
\section{Forecast for Einstein Telescope: Tests of General Relativity}
    After validating our method, we proceed to forecast how precisely ET can constrain deviations in the PN coefficients using BBH observations. For this analysis, we set the fiducial expansion point of the Fisher approximation to the PN coefficient $\phi_p$ predicted by GR, i.e., $\delta\phi_p^0 = 0$, for all individual observations.  

Given that the final ET design is not decided yet, we consider and compare three different detector configurations: a single 10km triangular configuration $\Delta$, a network of two L-shaped 15km misaligned detectors \texttt{2L\_45}, and a network of two L-shaped 15km aligned detectors \texttt{2L\_0}.
Further details on the detector setup, including the employed PSD, are provided in Appendix \ref{app:catalog_detector}. 

Given our observations in the previous section, we restrict to the \wfHM~waveform template. Thanks to its increased sensitivity, ET should more effectively exploit the physical information encoded in higher-order modes.
For the ET analysis, we use a lower frequency cutoff $f_\mathrm{min} = 2\mathrm{Hz}$. For a discussion on the impact of the choice for $f_\mathrm{min}$ see Appendix \ref{app:f_min}. 
Given ET's improved sensitivity, the number of events observed is expected to increase significantly. As a benchmark, during an observation run of about 3 to 4 months, we expect to have $\mathcal{O}(10^4)$ BBH sources \cite{Iacovelli:2022For, Begnoni:2025oyd}. 
Hence, for our forecast, we generated a catalog of $10^4$ BBH events, and by calculating the SNR and Fisher matrix, we selected the number of observable events among these. 
In this analysis, we drop the criterion based on the inspiral portion of the SNR, employed in Section \ref{sec:verification_with_lvk}. We verified that its inclusion does not affect the obtained results, suggesting that signals detected by ET generally show a sufficiently large inspiral SNR to robustly support this class of inspiral-based tests. Hence we define an event as observable if it satisfies the following two criteria: i) The event has a network SNR larger than 12, ii) the Fisher information matrix of the event is invertible. Note that the subset and number of observed events, selected from these 10,000 events in the catalog, depend on the detector configuration. For the details on the catalog, see Appendix~\ref{app:catalog_detector}.
In particular, $70.7_{-0.8}^{+0.7}\%$ of events in the catalog for the \texttt{$\Delta$}-configuration, $82.8_{-0.6}^{+0.6}\%$ for the \texttt{2L\_45}-configuration and $78.0_{-0.7}^{+0.6}\%$ for the \texttt{2L\_0}-configuration were used to compute the hierarchical posterior. We find that only very few events, of order $\sim1\%$, have been neglected based solely on a non-invertible Fisher matrix.
We emphasize that this procedure differs from the one in Section~\ref{sec:verification_with_lvk} and corresponds to calculating the constraints based on a fixed window of observation time, rather than a fixed number of observed events. However, this allows us to draw meaningful conclusions when comparing different detector designs, as done in this section. 

\begin{figure}
    \centering
    \includegraphics[width=1.0\textwidth]{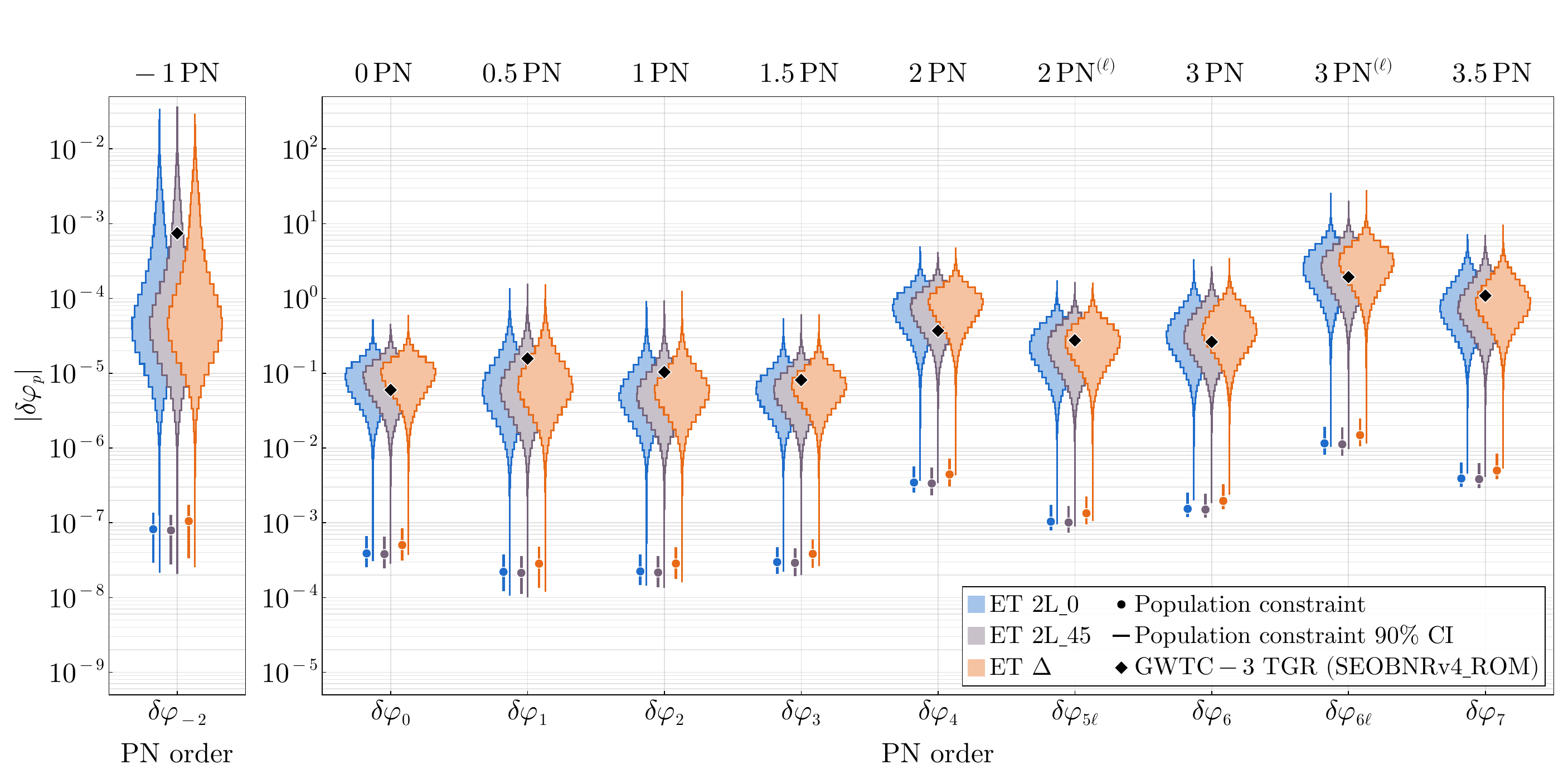}
    \caption{
    Forecast for the $90\%$ upper bounds on the magnitude of the post-Newtonian inspiral deviation coefficients $\delta\varphi_p$ for the three different ET configurations considered: triangular $\Delta$, two L-shaped interferometers with 45$^\circ$ misalignment \texttt{2L\_45}, two aligned L-shaped interferometers \texttt{2L\_0}. 
    The population constraints are obtained from the distribution $\PCc{\delta\varphi_p }{D^{\nobs}, I}$ in \eqref{eq:conditioned_deviation_posterior}, assuming that deviations take on the same value for all observed events, and are represented by three different markers, respectively. 
    The number of observed events used for the estimates correspond to a fixed observation period of approximately 4 months. 
    Each marker represents the median upper bound, obtained from $10^3$ independent realizations of the simulated observations, while the error bars represent the symmetric $90\%$ credible intervals. 
    The color-coded violin plots show the distribution of the single event $90\%$ upper bounds $|\delta\varphi_p|$.
    The black diamonds represent the LVK constraints from GWTC-3 \cite{LIGOScientific:2021test}.
    } 
    \label{fig:upper_bounds_deformation_coefficients_ET}
\end{figure}

Since we are performing the forecast for a GR consistency test, hence assuming GR as the null-hypothesis, we compute the combined $90\%$ credible upper bound $|\delta\varphi_p|$ from the conditioned deviation distribution $\PCc{\delta\varphi_p}{D^{\nobs}, I}$ in \eqref{eq:conditioned_deviation_posterior}. 
In appendix~\ref{app:a3_more_hyperparam_dist}, we further provide plots of the hierarchical hyperparameter distribution $\Pc{\mu,\sigma}{D^{\nobs}, I} $ in \eqref{eq:posterior_hyperparamter_tiger} and estimates from the hierarchical deviation distribution $\Pc{\delta\varphi_p}{D^{\nobs}, I} $ in \eqref{eq:deviation_dist_result}. The discussion of these results is deferred to the appendix since the main conclusions and most optimistic constraints on the improvements with ET can be inferred by using $\PCc{\delta\varphi_p}{D^{\nobs}, I}$.  
To quantify the statistical uncertainty of our forecast, due to specific catalog and detector noise realizations, we apply a bootstrap-type analysis \cite{Pacilio:2023uef}.
We generate a catalog of $2\cdot 10^5$ events from which we sample $10^3$ independent smaller catalog realizations, each containing $10^4$ events.
We use these to repeat the analysis a thousand times.
We depict the results of our forecast of the $90\%$ credible upper bound $|\delta\varphi_p|$ of GR deviations for ET in Figure~\ref{fig:upper_bounds_deformation_coefficients_ET}.
The marker in Figure~\ref{fig:upper_bounds_deformation_coefficients_ET} represents the median of $10^3$ realizations, while the error bars mark the symmetric $90\%$ credible interval, quantifying the statistical fluctuation of our forecast.  

\begin{figure}[t!]
\centering
\includegraphics[width=1.0\textwidth]{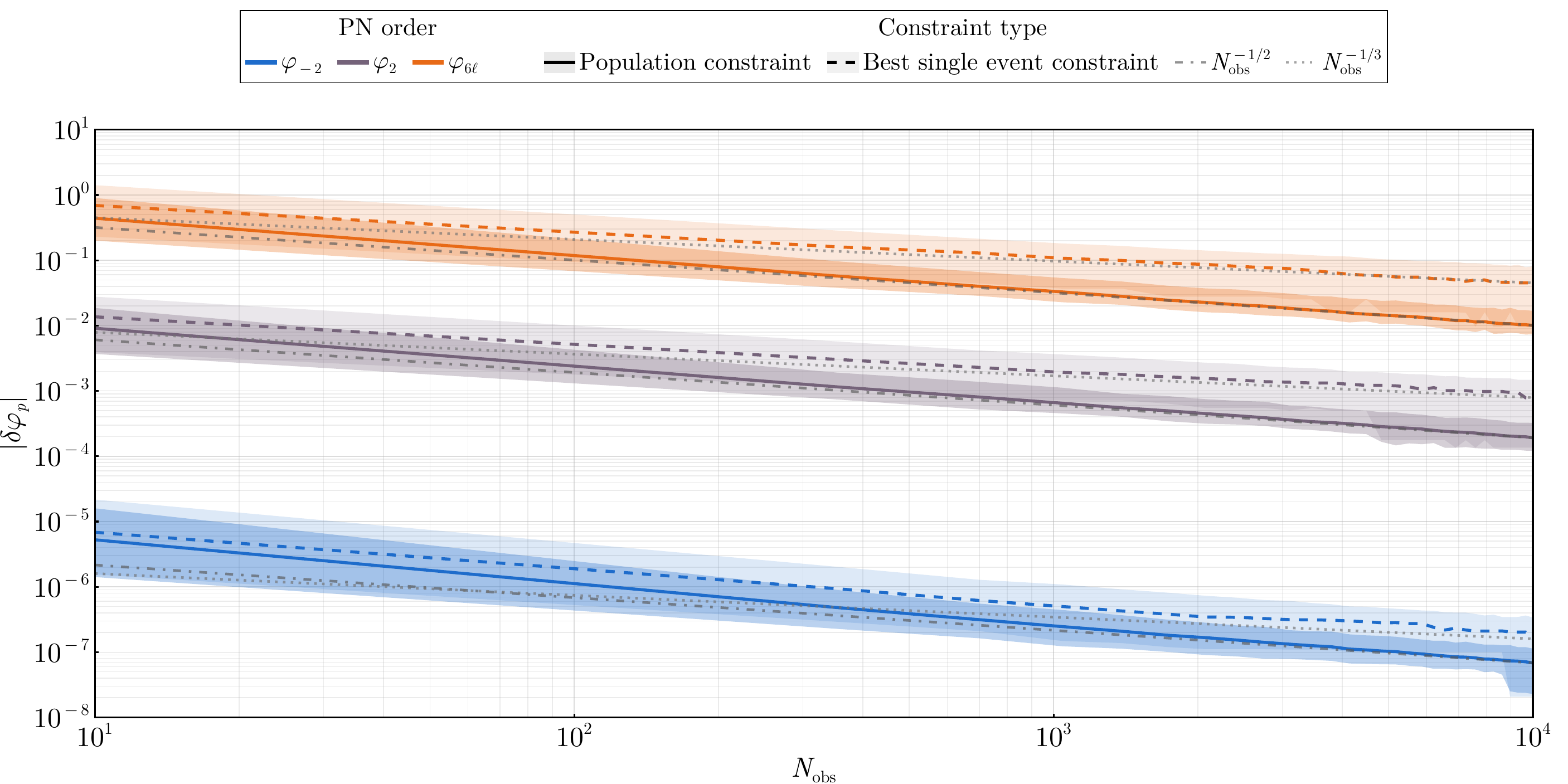}
\caption{Scaling behavior of the $90\%$ credible upper bounds on the magnitude of the post-Newtonian inspiral deformation coefficients as a function of the number of observed events $\nobs$. The solid lines show the median population upper bounds on $|\delta\varphi_p|$, obtained from the distribution $\PCc{\delta\varphi_p }{D^{\nobs}, I}$ in \eqref{eq:conditioned_deviation_posterior}. The dashed lines show the median of the tightest single event upper bounds $|\delta\varphi_p|$, among $\nobs$ events drawn from the catalog.
The median is calculated from various catalogs and noise realizations, with the faint color bands representing the resulting $90\%$ credible intervals. 
The dash-dotted and dotted lines represent different scaling limits, as expected respectively from the central limit theorem and the universal distribution of signal-to-noise ratio.
The detector was fixed to ET in the \texttt{2L\_45} configuration.}
\label{fig:upper_bounds_deformation_coefficients_ET_vs_N_events}
\end{figure}

The combined $90\%$ credible upper bounds from ET for BBH events, after an observation time of about four months, present an improvement between 2 to 4 orders of magnitude with respect to the GWTC-3 results \cite{LIGOScientific:2021test}, with larger improvements for lower PN orders. 
These improvements can be primarily attributed to three factors: ET's extended frequency range, down to $f_{\min} = 2\,\mathrm{Hz}$, its enhanced sensitivity across the entire frequency range, especially the region already accessible to the HLV network, and the increased statistical power from a larger number of detected events.
To estimate the improvement due to increased detector sensitivity, we plot the distributions of the estimated individual single event bounds, depicted as violin plots in Figure~\ref{fig:upper_bounds_deformation_coefficients_ET}. The largest improvement for single events occurs at lower PN orders, primarily due to the enhanced low-frequency sensitivity of ET. For $\delta\varphi_{-2}$, this results in an improvement of two orders of magnitude, while for $\delta\varphi_{0}$ through $\delta\varphi_{4}$ the bound is refined by about one order of magnitude. The contributions of the higher-order PN coefficients are entering only at higher frequencies and are less affected, see also Figure~\ref{fig:comparison_ET_GW150914_like_fmin} in Appendix~\ref{app:f_min}. 

To assess the improvements of the upper bounds due to the increased statistics, given the large number of expected detections, in Figure~\ref{fig:upper_bounds_deformation_coefficients_ET_vs_N_events} we plot the bounds as a function of the number of detected events, comparing the upper bounds obtained from the distribution $\PCc{\delta\varphi_p}{D^{\nobs}, I}$ to the tightest  $90\%$ credible upper bounds obtained from a single golden event, as a function of $\nobs$ observed events within a catalog realizations. 
For clarity, we fix the \texttt{2L\_45} ET detector configuration and show only the $-1$PN, $1$PN, and $3$PN log terms. 

We observe that for all PN orders the combined constraint yields tighter bounds with respect to the single event ones, with the relative gain increasing with the number of observed events $N_{obs}$, as expected. 
We find that for $N_{obs} = 10^4$, the average relative improvement between the two bounds is of $\mathcal{O}(3 - 5)$. The combined bounds, obtained from $\PCc{\delta\varphi_p}{D^{\nobs}, I}$, will further improve with more observed events.
The asymptotic scaling of the population constraints can be asymptotically approximated using the central limit theorem, i.e., $|\delta\varphi_p| \propto N_{obs}^{-1/2}$.
The scaling of the single events bounds, on the other hand, can be approximated asymptotically using the probability of observing a close-by high SNR golden event \cite{Chen:2014yla, Schutz:2011tw}, which results in $|\delta\varphi_p|_{\mathrm{golden}} \propto N_{obs}^{-1/3}$. 
The corresponding asymptotic improvement factor of the combined bounds over the golden-event bounds is therefore 
$|\delta\varphi_p|_{\mathrm{golden}}/|\delta\varphi_p| \propto N_{\mathrm{obs}}^{1/6}$; up to finite-$N_{\mathrm{obs}}$ corrections, the observed $\mathcal{O}(3\text{--}5)$ gain at $N_{\mathrm{obs}}=10^4$ is consistent with this expectation.
The faint color bands, quantifying the variance of the upper bound, due to different catalog and noise realizations, are narrower for the population bounds than for the best single event constraints. 
Considering the scaling of the bounds on the PN deviation coefficients with observation time, $|\delta\varphi_p|(t_{\mathrm{obs}}) \propto t_{\mathrm{obs}}^{-1/2}$, after $\mathcal{O}(years)$ of observations the bound on the deviation coefficient at 0PN order ($\delta\varphi_{0}$) from ET should be comparable to those from double pulsars \cite{LIGOScientific:2018dkp, Arun:2012hf, Barausse:2016eii, Lyne:2004cj, Kramer:2006nb, Yunes:2010qb, Wex:2014nva, Kramer:2016kwa, Shao:2017gwu, Freire:2012mg, Anderson:2019eay, Nair:2020ggs, Kramer:2021jcw, LIGOScientific:2026fcf}. To accurately compare the ET and the double pulsars bounds on the $-1$PN order ($\delta\varphi_{-2}$), 
the present analysis should be extended to also include binary neutron star systems \cite{LIGOScientific:2018dkp, Zhao:2021bjw, Chamberlain:2017fjl}. 

Regarding the constraints from ET, the three configurations considered yield quite similar and consistent results, with the \texttt{2L\_45} setup showing some slightly tighter bounds.  The observed differences among them remain well within the expected statistical scatter due to catalog realizations and noise fluctuations. This overall similarity could be an indication that the estimated BGR deviations, being related to intrinsic parameters, are quite insensitive to the specific detector geometry, as also noted in other papers (e.g., \cite{Iacovelli:2022For, Begnoni:2025oyd}).
As a side note, the presence of correlated noise between the detectors in the triangular configuration can have a non-negligible impact that will be important to quantify~\cite{Cireddu:2023ssf,Wong:2024hes,Caporali:2025mum}.
    \label{sec:forecast_et_gr_correct}
    
\section{Forecast for Einstein Telescope: Falsifying General Relativity}
    In the previous two sections, our analysis focused on forecasting consistency tests of GR. Thus, we set the fiducial point of the Fisher expansion to $\delta \varphi_p^0 = 0$ for all individual events and mostly used the conditional deviation distribution \eqref{eq:conditioned_deviation_posterior}. 
In the following, however, we assume General Relativity to not represent the true theory of gravity, and we are interested in forecasting the expected number of observed events, showing a small deviation away from GR, required to detect a significant deviation from GR.  

In general such deviations may manifest in several ways, possibly also depending on any of the compact binaries' parameters in non-trivial ways \cite{Payne:2024yhk, Bernard:2025dyh}. Due to the multitude of available theoretical models and the lack of experimental preference, we chose a minimal and self-consistent approach for our study.
To this end, we model the deviation parameter following a Gaussian population distribution, specified by two fiducial population hyperparameters $\mu^0_{\delta\varphi_p}$ and $\sigma^0_{\delta\varphi_p}$. 
The fiducial points for the individual Fisher approximated single event posteriors will then be drawn from a Normal distribution, i.e.
\begin{equation}
    \delta\varphi^{0}_p \sim \mathcal{N}\left(\mu_{\delta \varphi_p}^{0}, \sigma_{\delta \varphi_p}^{0}\right)\,.
\end{equation}
Hence, our analysis amounts to estimating the number of observed events required to recover the injected hyperparameter. 
We consider a deviation as significant if the fiducial deviation hyperparameters are recovered within the $90\%$ region of the hierarchical hyperparameter posterior, while the volume enclosed by the iso-surface containing GR is larger than $90\%$ \cite{Ghosh_2018}. This guarantees an exclusion of GR at the $90\%$ level.  
For this analysis, we utilize the full hierarchical hyperparameter distribution discussed in \cref{sec:hierarchical_analysis}.

\begin{figure}
    \centering
    \includegraphics[width=0.5\linewidth]{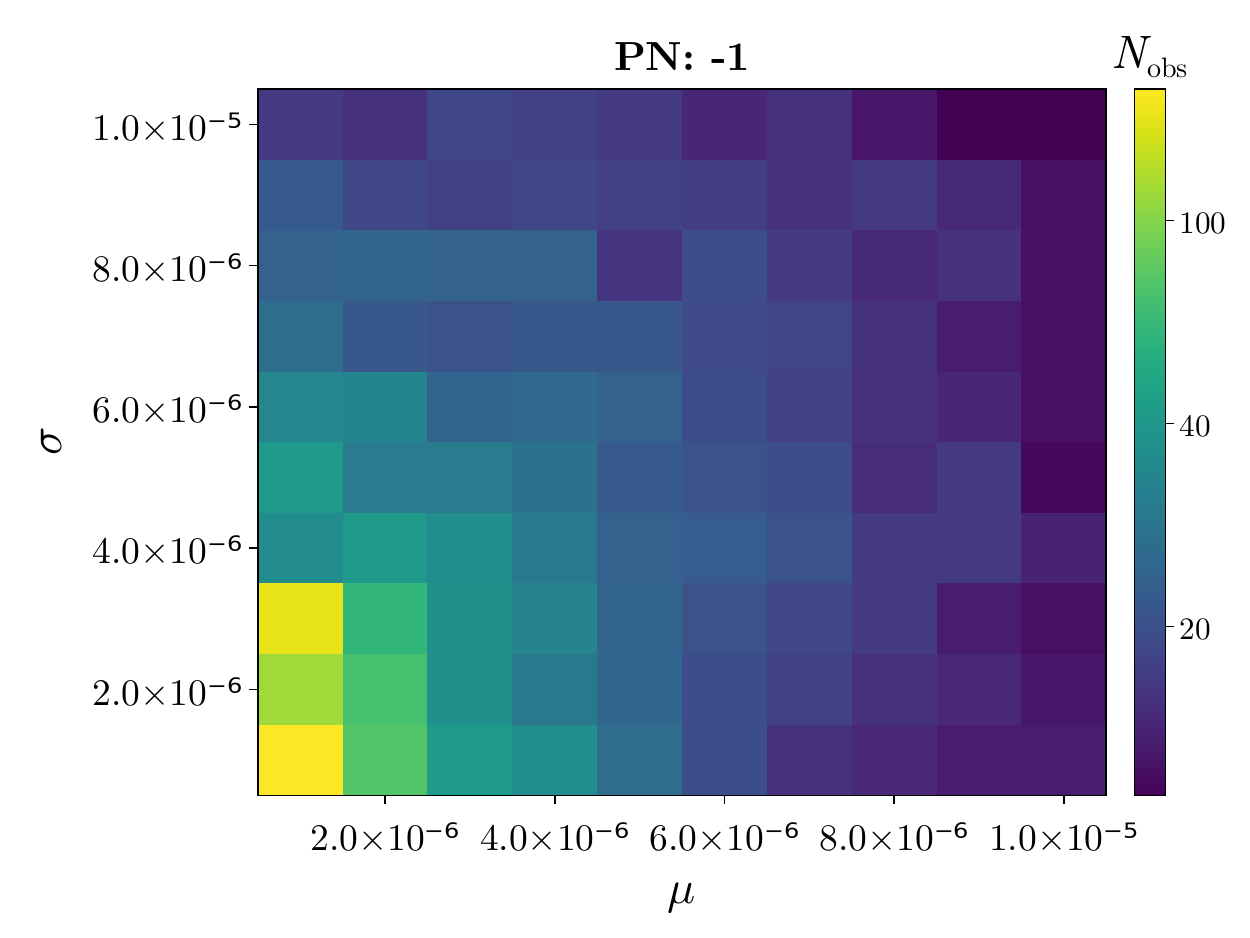}
    \includegraphics[width=0.49\linewidth]{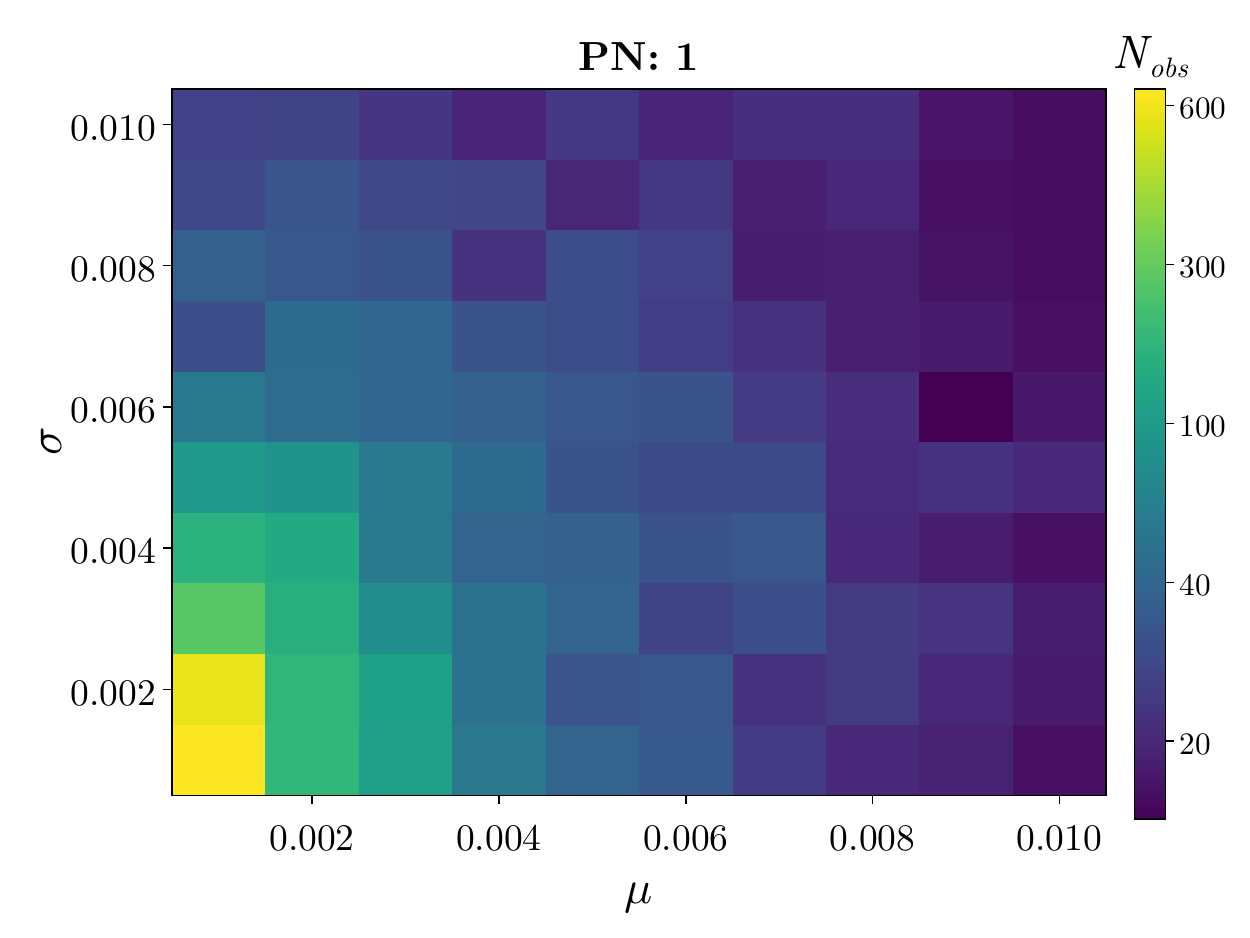}
    \caption{Number of detections needed to claim a deviation from GR with 90\% confidence level, given fiducial deviation parameter $\delta\varphi^0_p$, drawn from a Gaussian with mean $\mu_{\delta \varphi_p}^{0}$ and standard deviation $\sigma_{\delta \varphi_p}^{0}$. We plot the results for $\delta\varphi_{-2}$ on the left and for $\delta\varphi_2$ on the right, for the \texttt{2L\_45} configuration. The numbers plotted are the median of 150 repetitions of the forecast, with different catalogs and noise realizations.
    }
    \label{fig:grid_plot}
\end{figure}

In Figure~\ref{fig:grid_plot}, we show the number of detected events required to claim such a $90\%$ CL deviation of GR injecting fiducial hyperparameters $(\mu_{\delta \varphi_p}^{0}, \sigma_{\delta \varphi_p}^{0})$ for a fixed PN order and \texttt{2L\_45} as detector configuration. 
To estimate the minimum number of events required to claim a deviation from GR, we employ a bisection algorithm, repeating the hierarchical evaluation for different numbers of events.
Notably, such an estimate is very sensitive to the catalog and noise realization. Furthermore, individual golden events providing particularly tight constraints on deviations from GR may dominate the overall estimate. Thus, to mitigate these statistical fluctuations, we produce 150 catalog realizations and plot the median number of events required to claim the 90\% deviation from GR. 
From Figure~\ref{fig:grid_plot} we see the significant improvements that ET will bring, being able to constrain deviations of $\mathcal{O}(10^{-6})$ for $\delta\varphi_{-2}$ and $\mathcal{O}(10^{-3})$ for $\delta\varphi_{2}$ with only a few hundreds of observed events, i.e., within days of observations. 
This aligns with the results of Section~\ref{sec:forecast_et_gr_correct}: given the same number of events, the better sensitivity of ET will result in tighter bounds with respect to LVK. This effect is further compounded with the higher statistic due to the increased detection rate, such that few days of data taking will suffice to improve the LVK bounds.
In particular, considering the 1PN order, only $\mathcal{O}(20)$ events are required to exclude $(\mu_{\delta \varphi_p}^{0}, \sigma_{\delta \varphi_p}^{0}) = (0.01, 0.01)$ while $\mathcal{O}(600)$ events are required to exclude $(\mu_{\delta \varphi_p}^{0}, \sigma_{\delta \varphi_p}^{0}) = (0.001, 0.001)$, which corresponds to one week of data taking. 

    \label{sec:forecast_et_gr_wrong}
    
\section{Discussion and Conclusion}
    Testing General Relativity remains a primary objective of current and next-generation gravitational-wave detectors. Since the first detection of GW150914, and especially with GW170817, GW observations \cite{LIGOScientific:2016_150914, LIGOScientific:2017MM, LIGOScientific:2018mvr, LIGOScientific:2020ibl, KAGRA:2021TC3} have enabled a wide range of tests of GR, including measurements of the speed of GWs, constraints on spin-induced quadrupole moments, and bounds on the number of polarizations \cite{
LIGOScientific:2016lio, LIGOScientific:2018dkp, LIGOScientific:2019fpa, LIGOScientific:2017ycc, LIGOScientific:2021test, LIGOScientific:2020tif}. 

In this work, we address, for the case of ET, a central objective of the tests of GR: evaluating the consistency of the inspiral phase of compact binary coalescence with GR predictions. Modeling the inspiral via a PN expansion, we employ the TIGER framework to forecast the ET’s sensitivity to deviations in the PN coefficients.

We employed a Fisher-matrix approach, implemented with the \gwjulia code. First, to validate our framework, we compared our predictions with the GWTC-3 results reported by the LVK Collaboration, finding very good agreement with the published data (see Figure~\ref{fig:upper_bounds_deformation_coefficients_LVK}).
Then, we performed forecast analyses for ET, estimating the value of future bounds on potential deviations from GR, with the fiducial point of the Fisher approximation given by the PN coefficients, assuming GR as the fiducial model.

We analyzed a catalog of ten thousand binary black hole mergers, corresponding to roughly three to four months of observations, for each of the ten PN terms, considering three different ET configurations.
The analysis employed \texttt{IMRPhenomHM} waveform model, which includes higher modes. Alternatively, it would have been possible to use the effective-one-body (EOB) waveforms \cite{Cotesta:2018fcv, Pompili:2023tna} within the FTI framework \cite{Mehta:2022pcn,LIGOScientific:2021test,Sanger:2024axs}; however, such a choice is expected to have only a limited impact on our results \cite{LIGOScientific:2021test,Roy:2025gzv}.
We find that, after a few months of observation, each ET configuration results in tighter constraints on all PN coefficients by two to three orders of magnitude relative to current bounds, with the notable exception of the $-1$PN term, for which the improvement reaches four orders of magnitude. 
This significant improvement stems from the higher sensitivity of ET at lower frequencies, which enables more precise measurements of the early inspiral phase. This, in turn, underscores the importance of mitigating seismic noise and advancing cryogenic technology~\cite{Branchesi:2023COBA, ET:2025xjr}.
Furthermore, several years of observations should result in ET upper bounds for the 0PN order that could be competitive with double pulsars constraints.

Our analysis indicates that the several detector configuration considered all provide comparable bounds. The \texttt{2L\_45} design provides slightly better bounds across all PN orders, yet the difference is smaller than the variance due to catalog and detector noise realization. 

Furthermore, we carried out a theory-agnostic test to assess the constraining capabilities of ET in the presence of real deviations from GR. By injecting GR deviations with magnitudes drawn from a Gaussian distribution, we quantified ET’s ability to recover such deviations. Our results indicate that even a few days of observations would already exclude large regions of the beyond-GR parameter space. 

The present analysis highlights the scientific potential of ET, demonstrating its capability to provide precise tests of GR, and to significantly tighten current constraints when implementing the hierarchical Bayesian method. 
The efficiency and flexibility of our implementation, enabled by the \gwjulia code, were key to achieving the forecast results presented here. 
Our analysis also presents some approximations and limitations. While the Fisher matrix proved reasonably accurate in our validation, a full Bayesian study with more complete waveform models~\cite{Roy:2025gzv,Pompili:2023tna,Pompili:2025cdc,Ramos-Buades:2023ehm, vandeMeent:2023ols, Gamboa:2024imd, Albanesi:2025txj, Estelles:2025zah, Pratten:2020XAS, Pratten:2020ceb, Garcia-Quiros:2020qpx, Estelles:2021gvs, Planas:2025feq, Hamilton:2025xru} would provide more faithful estimates at the expense of high computational costs, which makes an analysis like this infeasible for now with a full Bayesian study. The source catalog could be extended by including binary neutron stars and neutron star-black hole binaries. In this work we assumed quasi-circular, non-precessing waveforms. A quantitative assessment of the impact and the potential for bias in the analysis \cite{Bhat:2024hyb, Krishnendu:2021fga, Saini:2023wdk, Narayan:2023vhm} arising from orbital eccentricity and spin-induced precession requires dedicated injection-recovery analyses with more complete waveform models, and this will be investigated in future work.
Moreover, future work should assess the impact of non-stationary or non-Gaussian detector noise, glitches, overlapping signals, waveform systematics, possible correlations between the GR deviations and astrophysical parameters \cite{Gupta:2024gun, Dhani:2024jja,Garg:2024qxq,Payne:2023kwj}, as well as correlated noise in the triangular configuration \cite{Cireddu:2023ssf, Caporali:2025mum}. 
If not properly mitigated, these systematic effects could fundamentally limit the accuracy of any future tests of General Relativity.

Other promising directions for future work include developing a more refined parametrization of beyond-GR effects. For instance, one could introduce explicit dependencies of the post-Newtonian deviations on binary parameters such as the mass ratio or total mass, or apply principal component analysis to capture correlated deviations more effectively \cite{Payne:2024yhk, Bernard:2025dyh, Mahapatra:2025cwk}.

Further extensions may involve testing broader classes of modified gravity scenarios, such as theories with massive gravitons \cite{Will:1997bb, Johnston:2025qmo, deRham:2014zqa}.
A natural next step is to apply this framework to binary neutron star or black hole–neutron star systems, where the higher event rates \cite{Begnoni:2025oyd, KAGRA:2021duu} and longer inspiral durations can provide tighter constraints, particularly at the low PN orders \cite{LIGOScientific:2018dkp}. Overall, the framework introduced here is highly flexible and can be straightforwardly extended to probe more complex and theory-specific deviations from GR. It thus provides a robust and computationally efficient foundation for forecasting tests of general relativity across diverse detector networks and astrophysical populations.

    \label{sec:conclusion}

\section*{Acknowledgments}
    We want to thank Gabriele Astorino, Nicola Bartolo, Alessandra Buonanno, Gabriele Demasi, Elisa Maggio, Pierpaolo Mastrolia, Lorenzo Pompili, Elise M. {S{\"a}nger}, Jan Steinhoff and Sebastian {V{\"o}lkel} for useful comments and insightful discussions. 
We thank Soumen Roy and Jan Steinhoff for valuable comments and feedback on the manuscript. We thank Yanbei Chen for orchestrating the LVK internal review. 
AB is supported by ICSC - Centro Nazionale di Ricerca in High Performance Computing, Big Data and Quantum Computing, funded by European Union - NextGenerationEU.
The work of M.P. is supported by the European Union under the Next Generation EU programme. M.P. gratefully acknowledges financial support from Fondazione Ing. Aldo Gini, from the INFN initiatives \textit{Amplitudes} and \textit{InDark}, and the hospitality and support of the Albert Einstein Institute. The PhD fellowship of J.P. is funded by the Italian Ministry of University through the project `Nano-Meta-Materials and Devices: New Frontier Concepts for Particle and Radiation Detection' (Grant `Dipartimento di Eccellenza' 2023-2027, CUP I57G22000720004) at the Department of Physics of the University of Pisa.
CloudVeneto is acknowledged for the use of computing and storage facilities.
This material is based upon work supported by NSF's LIGO Laboratory which is a major facility fully funded by the National Science Foundation.

\section*{Data/code availability statement}
    All the data and plots presented in this work can be reproduced with the publicly available code provided at \cite{code_base}.

\pagebreak
\appendix

\section{Catalog and detector setup}
\begin{figure}[t!]
    \centering
    \includegraphics[width=0.95\linewidth]{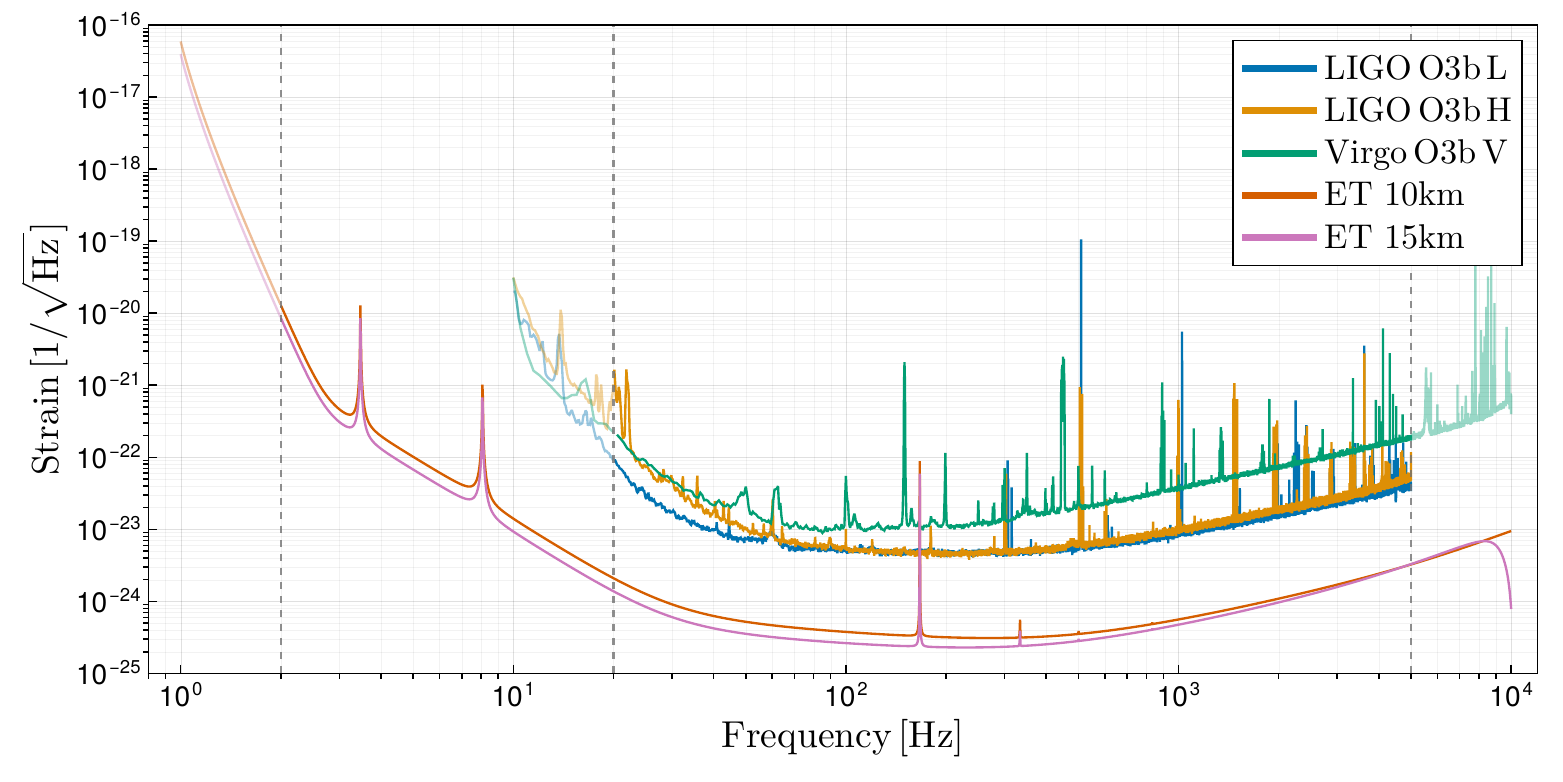}
    \caption{Plot of the detector amplitude spectral densities (ASD) for detectors used in this work: $\mathtt{H}$, $\mathtt{L}$, and $\mathtt{V}$ during the O3b observation run and ET with $10$km and $15$km arm length. The faint parts of the ASDs mark the regions below $f_\mathrm{min}$ and above $f_\mathrm{max}$, omitted in the Fisher analysis. }
    \label{fig:sensitivity-curves}
\end{figure}

    This appendix provides more details on some important quantities for our analysis. In particular, we introduce and characterize the properties of the detectors and we comment on the catalog properties.
Each detector is determined by its shape, position on Earth, orientation, and the PSD. The shape, i.e., the geometry, is defined through the opening angle between the arms $\zeta$, which is $90^{\circ} $ for L-shaped and $60^{\circ} $ for triangular detectors. The latitude $\lambda$ and longitude $\varphi$ define the position of the center of the detector, i.e., the location of the beam-splitter for L-shaped and the center of the triangle for triangular detectors. Then, the configuration is completely described with the orientation angle, representing a rotation in the detector plane, which is particularly important for a network of detectors.
Each detector has an orientation $\gamma$ w.r.t. a given direction, which in this work (and in \gwjulia) we choose w.r.t. the local East.
The current network proposals for ET contemplate~\cite{Branchesi:2023COBA, ET:2025xjr} a triangular design or 2L networks, and, in the latter case, there is also the freedom of the orientation between the two detectors, $\beta$. The optimal configuration to measure CBC corresponds to two detectors misaligned with an angle of $\beta_{\rm CBC} = \pi/4 + n\pi/2$, where $n$ is an integer\footnote{Notice that this statement is only valid statistically, i.e., when dealing with a large number of sources. Indeed, some CBCs may be better measured with a different orientation angle between the two detectors. }.  Once the curvature of the Earth is considered, the detectors are not coplanar, so the orientation between the two detectors is not $\beta$~\cite{Flanagan:1993ori, Christensen:1996origwb} and we define it as $\alpha = \gamma_2-\gamma_1$. This way, we obtain $\alpha= \beta+2.51^{\circ}$ for two detectors, one in Sardinia, one in the MR region. All the following considerations apply to $\beta$, which is the physical parameter in the FIM evaluation.
Finally, the PSDs used for the comparison with GWTC-3 \cite{KAGRA:2021TC3, LIGOScientific:2021test} are the ones corresponding to the O3b run\footnote{from \cite{ligo_scientific_collaboration_and_virgo_2023_7997424},
the data that was used to create Figure 2 in the GWTC-3 analysis.}  \cite{aLIGO:2020wna, Tse:2019wcy, Virgo:2019juy,Virgo:2022ysc} and in \cref{fig:sensitivity-curves} we show the amplitude spectral densities, i.e., the square root of the PSD.

Since ET's final design and configuration have not been decided yet~\cite{Branchesi:2023COBA, ET:2025xjr}, we consider different setups to compare their capabilities in detecting and estimating the parameters of CBCs. In particular, we consider: 
    \begin{itemize}
        \item $\Delta$: triangular ET with 10 km arms in Sardinia
        \item \texttt{2L\_0}: two 15 km L-shape interferometers, one in Sardinia, one in the MR Euroregion. The orientations of the two detectors are chosen such that $\beta=0^\circ$.
        \item \texttt{2L\_45}: same as \texttt{2L\_0} with the exception that the orientations lead to $\beta=45^\circ$. This is expected to be the best network for measuring BGR parameters.
    \end{itemize}
All the ET networks have cryogenic technology, improving the sensitivity to lower frequencies. 
We then consider two PSDs \cite{Branchesi:2023COBA, Hild:2010id}, one for a detector arm length of 10km for the triangular design and one of 15km arm length for the L-shape.
In this work, and in \gwjulia, we also consider that the beam splitters presented in the triangular configuration have slightly different positions.
We also include the effects caused by the Earth's rotation, which are supposed to help marginally the FIM analysis, in the case of BBH \cite{Begnoni:2025oyd}.

A CBC is identified by numerous parameters, both intrinsic, i.e., characterizing the source properties, and extrinsic, i.e., illustrating the relation between the detector and the source, which we collectively denote with $\boldsymbol{\theta}$. Among the intrinsics, in this work, we consider the masses $m_{1/2}$ and the longitudinal spins $\chi_{1/2\,z}$.
Within the extrinsic, we instead consider: the inclination angle $\iota$, i.e., the angle between the binary angular momentum and Earth position, the sky position, indicated with $(\theta,\phi)$, the polarization angle $\psi$ and the time and phase to coalescence, indicated respectively with  $t_{\rm coal}$ and $\Phi_{\rm coal}$.

When possible, we rely on the recent populations provided by the LVK collaboration~\cite{KAGRA:2021duu}. 
    In particular, the masses are drawn from a \texttt{PowerLaw + Peak} distribution~\cite{KAGRA:2021duu}, which consists of a power law plus a Gaussian peak at $M_{\rm peak}\approx 34 M_{\odot}$ and the second mass is drawn from a power-law proportional to $\propto q^{-1.1}$~\cite{KAGRA:2021duu}. 
    Regarding the spins, in the case of BBH, we employ the Default model from~\cite{KAGRA:2021duu}. For the redshift distribution, we use the Madau-Dickinson distribution~\cite{Madau:2014Dickinson}, which is convoluted with a time delay distribution. The time delay represents the time between the star formation and the merger of the BBH, and we consider a power-law distribution. A key and still uncertain parameter in this distribution is the minimum time delay allowed $t_{d,\rm{min}}$, which we consider 20 Myrs~\cite{Mapelli:2017hqk}.
    The sky positions are drawn from a uniform distribution over the sphere, i.e., $\varphi$, the azimuthal angle is drawn from a uniform distribution, and $\theta$, the polar angle, is uniform in the cosine. Similarly to $\theta$, the inclination angle $\iota$ is drawn from a uniform distribution in the cosine.
    Then, the polarization angle $\psi$ represents the rotation needed to align the source frame, where the $h_{+/\times}$ polarizations are defined, with the detector frame~\cite{whelan2013geometry}. This angle is sampled uniformly in $[0,\pi]$.
    The time to coalescence, $t_{\rm coal}$, represents the time of the merger and is uniform in a day. The phase to coalescence $\Phi_{\rm coal}$ represents the phase at which the merger happens and, similar to $t_{\rm coal}$, it is drawn from a uniform distribution, this time in the interval $[0,2\pi]$.
    
    Finally, for the values of local merger rate, i.e., the number of sources in a volume at redshift zero, we consider $R_0 = 17\, [\text{Gpc}^{-3} \text{yr}^{-1}]$~\cite{KAGRA:2021duu}.  Moreover, the total number of sources in one year, which is obtained by integrating the redshift distribution of the sources, after the convolution with the time delay distribution, times the local rate, leads to
    $n_{\rm sources} = 3.35 \times 10^4\,\text{yr}^{-1}$.
    In this work, we do not consider the overlap of different waveforms in the detector, 
    which should be considered in a more refined analysis.

    \label{app:catalog_detector}
    
\section{The impact of the low frequency cutoff $f_\mathrm{min}$}
    \begin{figure}[t!]
    \centering
    \includegraphics[width=1\textwidth]{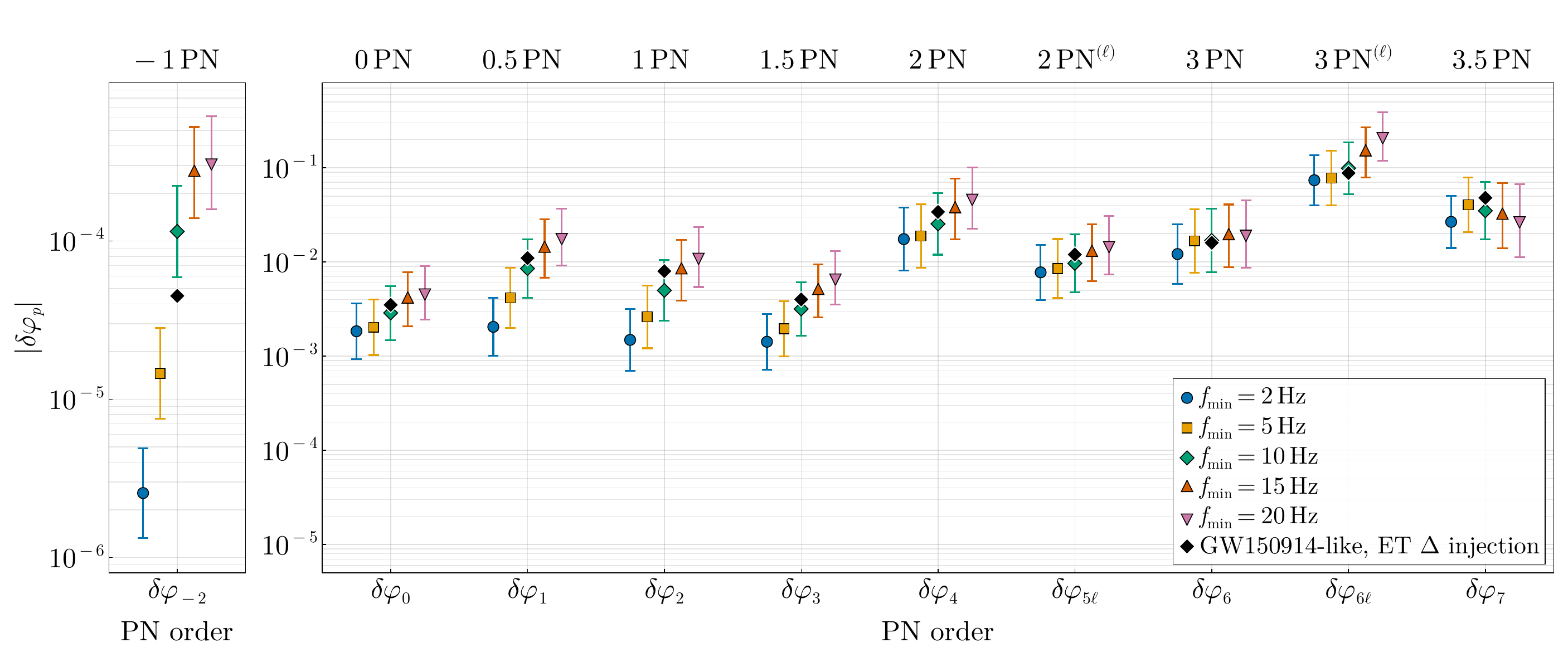}
  
    \caption{Forecast of the 90\% upper bounds on the magnitude of the post-Newtonian deformation coefficients $\delta \varphi_p$ as a function of the minimum frequency $f_{\mathrm{min}}$ used in the analysis, for a GW150914-like event. The forecast was performed using the \wfHM~waveform model for the Einstein Telescope \texttt{$\Delta$} configuration by varying the minimum frequency $f_{\mathrm{min}}$ of the signal, as reported in the legend. 
    The markers and the error bars for several $f_{\mathrm{min}}$ configurations quantify, respectively, the average value and the 90\% confidence interval of the bounds over 100 realizations of a GW150914-like event. This allows for taking into consideration detector noise realizations and different parameters of the event, keeping masses and luminosity distance fixed according to GW150914.
    We also report with dark blue diamonds the bounds of figure~1.1 of \cite{ET:2025xjr}, for comparison of our result with their Bayesian injection-recovery analysis of a comparable event in the ET \texttt{$\Delta$} configuration.
    }  
    \label{fig:comparison_ET_GW150914_like_fmin}
\end{figure}

This appendix discusses the choice of the minimum frequency used in the analysis. We explore the effect of changing $f_{\rm min}$ in the PSD in \cref{fig:sensitivity-curves}, in particular, $f_{\rm min}$ acts as a hard cut-off, setting the PSD infinite at lower frequencies. 
In \cref{fig:comparison_ET_GW150914_like_fmin} we consider a GW150914-like event, for different values of $f_{\rm min}$ to asses the impact of the low frequency cutoff on the obtained GR deviation constraints at different PN orders in the case of ET. The error bars in the plot stem from sampling multiple GW150914-like events, with the binary component masses and the luminosity distance fixed, but varying the other binary parameters such as spins, inclination etc., as well as using different noise realizations for each event.
The influence of $f_{\rm min}$  is particularly important for the $-$1PN term, though other PN terms are also affected. As expected, lower-order PN terms are generally more sensitive to changes in $f_{\rm min}$, with the 0PN and 3PN terms being notable exceptions. For a direct comparison, \cref{fig:comparison_ET_GW150914_like_fmin} also includes data points from Figure 1.1 of \cite{ET:2025xjr}, where the authors conducted a full Bayesian parameter estimation of a comparable event. This allows for a clear comparison of our Fisher analysis with their full injection-recovery study.

    \label{app:f_min}
    
\section{Fisher approximated hyper-parameter distribution}
\begin{figure}[t!]
    \centering
    \includegraphics[width=0.97\linewidth]{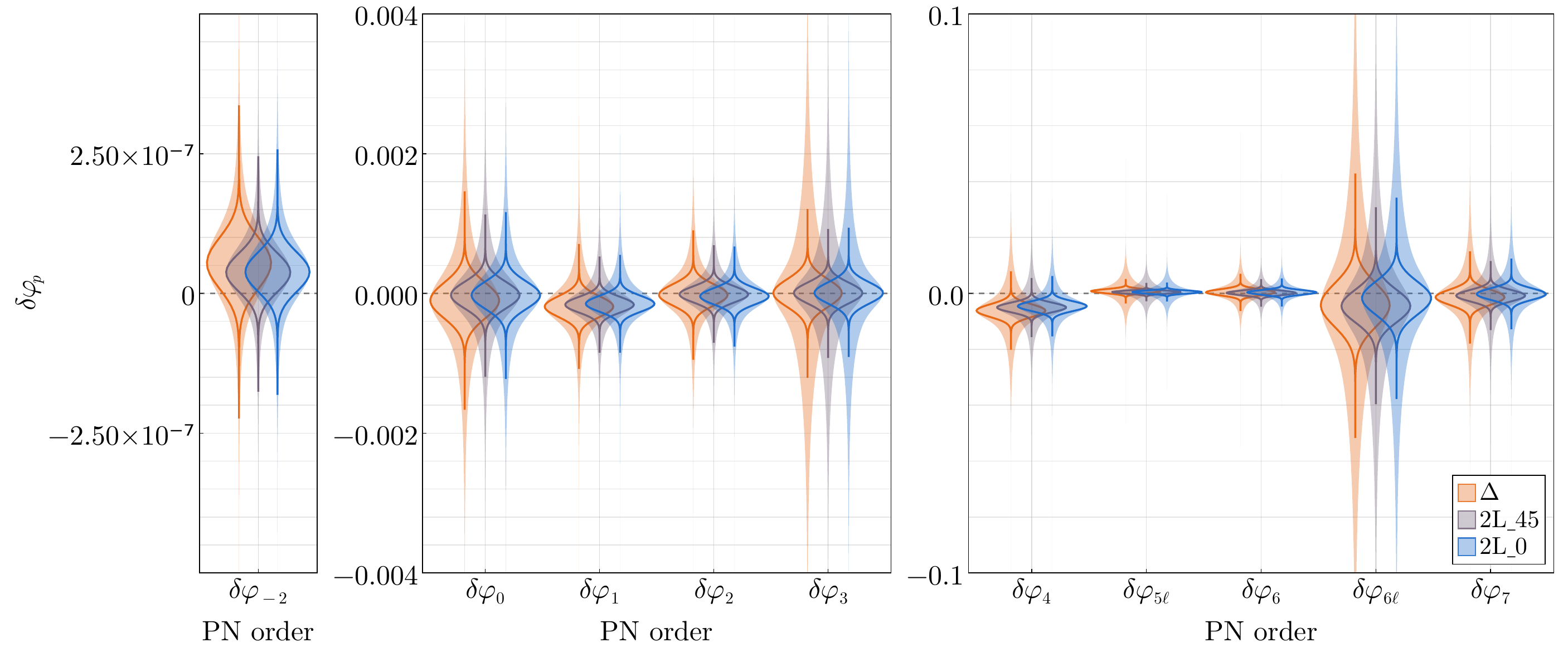}
    \caption{Visualization of the hierarchically combined posterior distribution \eqref{eq:deviation_dist_result} for the PN deviation coefficients $\delta\varphi_p$, as obtained from the Fisher approximation, for all three ET configurations considered in this work. 
    The distributions have been obtained by using a catalog of $10^4$ events corresponding to an observation period of approximately four months, as described in Section~\ref{sec:forecast_et_gr_correct}.
    The filled violin plots represent the posterior distribution $\Pc{\delta\varphi_p}{ D^{\nobs}, I}$ for the $\delta\varphi_p$ deformation coefficients. The unfilled violin plots, with thick outline, represent the conditioned posterior distribution $\PCc{\delta\varphi_p}{D^{\nobs}, I}$, describing the scenario in which all GR deviations manifest with the same value in all single events.    
    }
    \label{fig:posterior_delta_phi_main}
\end{figure}

    In this appendix, we show explicit results of the Fisher approximated hyperparameter distributions, comparing the full two-dimensional distributions to the conditioned ones. In Figure~\ref{fig:posterior_delta_phi_main} we depict the resulting distribution in the deviations $\delta \varphi_p$ for all PN orders, both for the generic result~\eqref{eq:deviation_dist_result}, and for the conditioned result~\eqref{eq:conditioned_deviation_posterior}, which assumes all the GR deviations take on the same value for all of the single events. As can be seen from Figure~\ref{fig:posterior_delta_phi_main}, the conditioned posterior is always narrower, reflecting the additional prior information.

\begin{figure}[t!]
    \centering
    \includegraphics[width=0.97\linewidth]{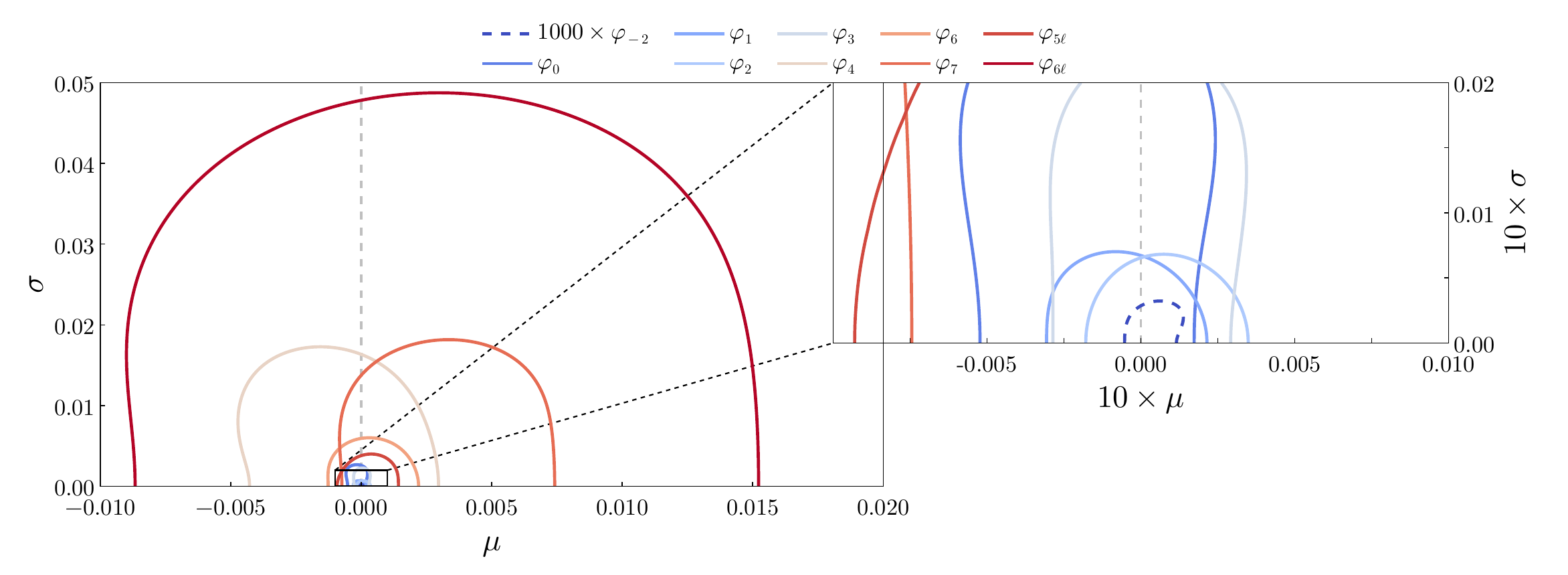}
    \caption{Visualization of the hierarchical hyperparameter distribution $P(\mu, \sigma | D^{\nobs}, I)$ for the \texttt{2L\_45} detector configuration of ET. 
    The contours show the 90\% credible region. The inset is scaled by a factor of 10. The contour plot corresponding to the $-1$PN order has been scaled by a factor of 1000.
    }
    \label{fig:posterior_mu_sigma_main}
\end{figure}

In Figure~\ref{fig:posterior_mu_sigma_main} we depict the $90\%$ contours, as obtained from the Fisher approximated two-dimensional hyperparameter distribution. For the analysis in Section~\ref{sec:forecast_et_gr_wrong}, these contours were used to decide whether GR is contained in the $90\%$ credibility region. 

Comparing these plots depicted in Figures~\ref{fig:posterior_delta_phi_main} and \ref{fig:posterior_mu_sigma_main} with the corresponding posterior distributions of the GWTC-3 analysis \cite{LIGOScientific:2021test}, we find a similar improvement as in Figure~\ref{fig:upper_bounds_deformation_coefficients_ET}, with distributions narrower between a factor of $\mathcal{O}(10^2)$ to $\mathcal{O}(10^4)$, especially for lower PN orders. 
    \label{app:a3_more_hyperparam_dist}

\section{On the role of higher-modes and selection effects for LVK}
    To clarify the origin of the tails in the distributions of single-event constraints for \wfHM, observed in the \texttt{HLV} network analysis, we examine how the constraints improve when transitioning from \wfD~to \wfHM~as a function of the binary parameters.
Since the improvements are most pronounced for $\varphi_7$, we will focus on the 3.5PN order. However, a similar discussion also holds for $\varphi_{5\ell}$  and $\varphi_6$. From the entire generated catalog, we restrict to the distribution of events with redshift $z<0.5$, since the largest portion of observable events lies within this redshift range. 
Further, we restrict to events with detector frame chirp mass $(1+z)\mathcal{M}<80 M_\odot$, covering almost all events defined as observable in section~\ref{sec:verification_with_lvk}.
\begin{figure}[t!]
    \centering
    \includegraphics[width=1.0\linewidth]{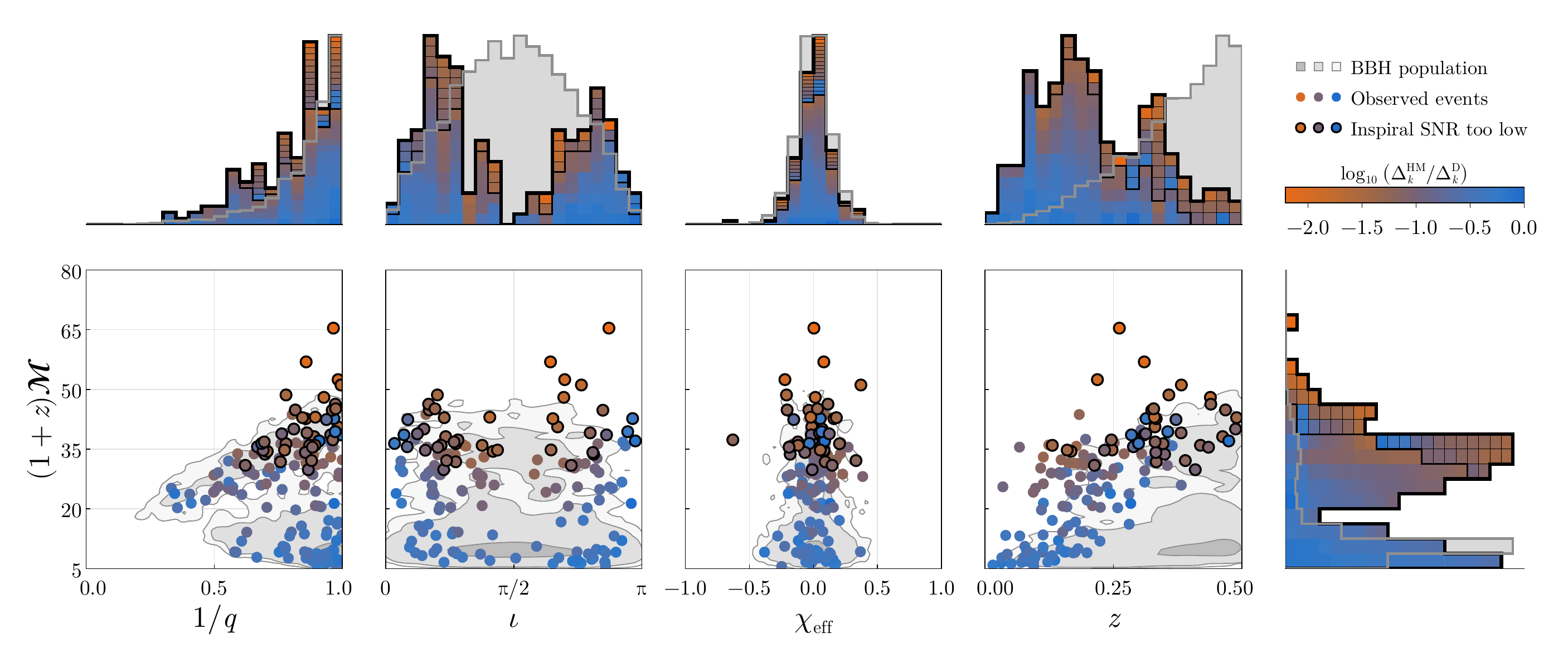}
    \caption{
        Visualization of the distribution of events for the single event constraints on $\delta\varphi_7$ for the \texttt{HLV} network. 
        The distribution of event parameters generated by the catalog is shown in grey.
        The colored histograms and scatter plots show events with SNR$>12$ and an invertible Fisher matrix.
        These events are further divided into two groups: observable events (inspiral SNR$\geq 6$) and events with insufficient inspiral SNR (inspiral SNR$<6$).
        The color encodes the improvement in the single event Fisher error $\Delta_k$, see \eqref{eq:delta_k_definition}, when comparing \wfD~to \wfHM.
        The subset of events which are omitted from the analysis in Section~\ref{sec:verification_with_lvk} due to their insufficient inspiral SNR are marked with black outlines in the scatter plot.
        In the histogram, each colored square represents one event, and events with insufficient inspiral SNR are marked with a black frame.
        }
    \label{fig:lvk_additional_checks}
\end{figure}

The histograms in Figure~\ref{fig:lvk_additional_checks} show a comparison between the distribution of single event parameters, before and after imposing the selection criteria that define an event as observable, and omitting the selection criteria on the inspiral $\mathrm{SNR}$. We see that after the selection, a bimodal structure in the chirp mass distribution emerges. The peak in the mass reflects the Gaussian peak at $M_\mathrm{peak} \approx 34\, M_\odot$ in the primary source mass distribution of the population model \cite{KAGRA:2021duu}, used to generate our catalogs, see Appendix~\ref{app:catalog_detector}.
Further, we notice that the improvements between the two waveforms increase with the chirp mass and are most prominent for inclination angles around $\iota \approx \frac{\pi}{4}$, being the region at which higher multipole contributions are most visible \cite{Puecher:2022sfm}.

High mass detected events generate most of their signal in the late-inspiral and merger part of the waveform,
making them non optimal candidates for BGR inspiral tests. Nonetheless, these high mass events contribute
significantly to the fraction of observed LVK events due to their large detection horizon and the peak in the mass distribution of the population model. 
As evident from Figure~\ref{fig:lvk_additional_checks}, the most pronounced differences between the two waveform models under consideration arise for the high-mass events. The improvement observed when transitioning from \wfD~to \wfHM~can be attributed to the enhanced contribution of the typically subdominant higher multipoles relative to the fundamental mode in these systems \cite{Puecher:2022sfm}.

The GR deviations have been parameterized by allowing modifications of the PN coefficients in the inspiral phase, which have been extended to the rest of the waveform by matching conditions, adjusting the onset-phase and time of the merger and ringdown. Moreover, we would like to stress that the magnitude of improvement has to be interpreted with caution, due to the short signal duration in the inspiral phase for these events. We thus consider the unexpectedly large improvements seen when using \wfHM~to be a potential concern that merits additional scrutiny.
This is further corroborated by looking at Figure~\ref{fig:lvk_additional_checks}, where events with inspiral $\mathrm{SNR}<6$ are marked by black outlines, providing the largest improvements when including higher modes. 
These events are excluded from the analysis resulting in Figure~\ref{fig:upper_bounds_deformation_coefficients_LVK}, by the definition of observable events for the HLV network, see Section~\ref{sec:verification_with_lvk}.
Due to the frequency dependence of the various PN orders, explicit in \eqref{eq:pn_inspiral_phase_formula}, the higher PN coefficients are more sensitive to this effect.
The combination of the bimodal structure in the chirp mass and the trend of improvements with increasing mass allows to explain the long tail in the single event distribution of $\delta\varphi_7$ observed in Figure~\ref{fig:upper_bounds_deformation_coefficients_LVK}.
It is noteworthy that for ET differences between the two waveforms are less pronounced across the whole parameter space, due to ET's lower minimum frequency $f_\mathrm{min} = 2\ \mathrm{Hz}$ and overall improved sensitivity, allowing to have access to more information from the earlier inspiral portion of the signal, especially for the higher mass events present in the population. For this reason, we could omit the inspiral criterion for the results presented in Section~\ref{sec:forecast_et_gr_correct}.

    \label{app:a4_lvk_additional_checks}
      
\bibliographystyle{JHEP}
\bibliography{bibliography.bib}

\end{document}